\documentclass[preprint,superscriptaddress,amsmath,amssymb,aps,physrev]{revtex4-1}
\usepackage{graphicx,dcolumn,hyperref,bm,amsmath,xcolor,amssymb,amsfonts,marginnote}
\usepackage[separate-uncertainty=true,multi-part-units=single]{siunitx}
\usepackage{subcaption}
\usepackage{cancel}

\usepackage{color}

\newcommand*\colvec[3][]{
	\begin{pmatrix}\ifx\relax#1\relax\else#1\\\fi#2\\#3\end{pmatrix}
}

\begin{document}

\title{Irreversible hydrodynamic trapping by surface rollers}

\author{Alexander Chamolly}
\email{ajc297@cam.ac.uk}
\affiliation{%
Department of Applied Mathematics and Theoretical Physics, University of Cambridge, Cambridge CB3 0WA, United Kingdom
}%
\author{Eric Lauga}%
\email{e.lauga@damtp.cam.ac.uk}
\affiliation{%
Department of Applied Mathematics and Theoretical Physics, University of Cambridge, Cambridge CB3 0WA, United Kingdom
}
\author{Soichiro Tottori}
\email{st607@cam.ac.uk}
\affiliation{Cavendish Laboratory, Department of Physics, University of Cambridge, CB3 0HE, United Kingdom}%

\date{\today}

\begin{abstract}
A colloidal particle driven by externally actuated rotation can self-propel  parallel to a rigid boundary by exploiting the hydrodynamic coupling that surfaces induce between translation and rotation. As such a roller moves along the boundary it generates local vortical flows, which can be used to trap and transport passive cargo particles. However, the details and conditions for this trapping mechanism have not yet been fully understood.
Here, we show that the trapping of cargo is accomplished through time-irreversible interactions between the cargo and the boundary, leading to its migration across streamlines into a steady flow vortex next to the roller. The trapping mechanism is explained analytically with a two dimensional model, investigated numerically in three dimensions for a wide range of parameters and is shown to be analogous to the deterministic lateral displacement (DLD) technique used in microfluidics for the separation of differently sized particles. The several geometrical parameters of the problem are analysed and we predict that thin, disc-like rollers offer the most favourable trapping conditions.
 
\end{abstract}

\maketitle


\begin{figure}
	\centering
	\includegraphics[width=.8\linewidth]{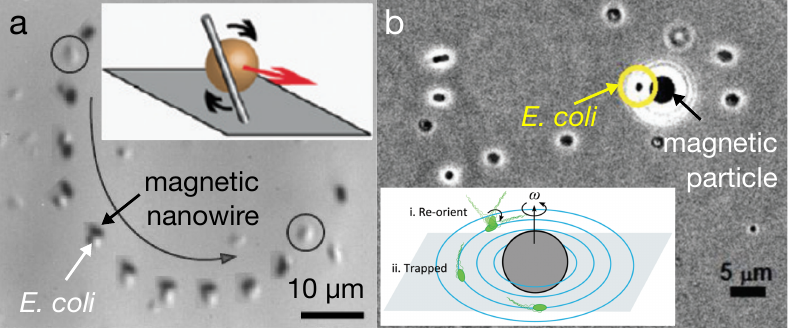}
	\caption{(a) A rotating magnetic nanowire traps and transports \textit{Escherichia coli} (\textit{E.~coli}). 			
	Rotating axis is parallel to the bottom surface.
	Adapted from Ref.~\cite{Petit2012} with permission.
	Copyright (2012) American Chemical Society.
	(b) A rotating magnetic microparticle traps and transports \textit{E.~coli}. 
	Rotating axis is slightly tilted ($15^\circ$) from the normal direction to the bottom surface.
	Adapted from Ref.~\cite{Ye2014} with permission from the Royal Society of Chemistry.
	\label{fig:Introduction}}
\end{figure}

\section{Introduction}
One of the first practical skills acquired by babies is the catching and moving of small items. The fluid world provides a similar challenge and the entrapment and manipulation of small objects has long been of great technological interest in micro- and nanofluidics. A wide range of physical mechanics may be exploited to achieve these tasks, giving rise to optical \cite{Grier2003}, magnetic~\cite{DeVlaminck2012}, electrostatic \cite{Krishnan2010}, or hydrodynamic forces \cite{Shenoy2016}.
In recent years, synthetic swimmers actuated by external fields, chemical fuels or bacteria have been attracting attention and been employed successfully for the transport of cargo towards biomedical applications~\cite{Kagan2011, Tottori2012, Gao2012, Alapan2018}. 

In what is perhaps the simplest configuration suitable for the manipulation of objects in a fluid at small scales, rotating  nanowires have been shown  to be capable of trapping and transporting small particles within hydrodynamic vortices, as shown in Fig.~\ref{fig:Introduction}a~\cite{Petit2012, Zhou2017, Mair2017}.
These magnetic nanowires are made of nickel and located near a flat  surface.
When the rotational axis of the magnetic field is parallel to that interface, asymmetric viscous drag near the wall converts rotation of the wire into a translation force in the direction parallel to the surface~\cite{Tierno2008, Sing2010}. 
When the nanowire then rotates near a non-magnetic body (in this particular case the bacterium \textit{Escherichia coli}), the body can be trapped and transported by the resulting vortical flow.

Similar results have also been reported using magnetic particles rotating along an axis slightly tilted from the perpendicular direction, as shown in Fig.~\ref{fig:Introduction}b~\cite{Ye2014}. Related hydrodynamic bound phenomena have been demonstrated using a pair of magnetically driven rollers~\cite{martinez2018emergent,Delmotte2018}.
In general, when a body rotates and translates simultaneously, a  vortical flow field of finite size appears around the body. 
Both fluid and particles in this region are transported together with a moving roller, as was previously shown for a rod simultaneously rotating and translating in an unbounded fluid~\cite{Zhao2018}.
However the onset of trapping demonstrated experimentally has so far remained elusive and both the physical mechanism behind the  trapping and   the optimal trapping conditions have yet to be identified.

At relatively high Reynolds number, inertial forces have been exploited to focus or trap particles in microfluidic channels~\cite{DiCarlo2009, Amini2014, Vigolo2014}. 
However, in the low Reynolds number limit that is relevant for small particles, inertial terms become negligible and the fluid motion is quasi-steady. 
Specifically, the fluid velocity, ${\bm u}$, satisfies the incompressible Stokes equations~\cite{Happel1983}
\begin{align}\label{eq:Stokes}
\nabla p = \mu \nabla^2 {\bm u},\quad  \nabla \cdot  {\bm u} = {\bm 0},
\end{align}
where $p$ and $\mu$ are the dynamic pressure and   viscosity of the fluid, respectively.
Since these equations have no explicit time dependence, no time-irreversible motion and thus no focusing and entrapment is possible unless irreversible forces are introduced through the boundary conditions.

In this article, we demonstrate that the mechanism of hydrodynamic trapping by a surface roller is due to the steric interaction of cargo particles with the solid boundary. Specifically, when the   cargo is advected by the flow created by the roller and also sufficiently large, the steric interactions with the bounding surface  allow it to migrate across streamlines into the steady flow vortex and it remains trapped there. We begin by investigating the mechanism numerically using a model roller and 
finite-element simulations that we describe in \S\ref{sec:compmod}. Our results are summarised in \S\ref{sec:comres}, where we present a phase diagram that indicates which parameter configurations lead to trapping and which do not. Furthermore, we illustrate the process of cargo migration and investigate the case of pure translation and no rolling. In \S\ref{sec:theory} we present two theoretical models focusing on different aspects of our setup and explaining different features of the phase diagram, as well as the physical mechanism of cargo trapping. The paper concludes with a discussion in section \S\ref{sec:discussion} where we show
in particular that trapping is analogous to deterministic lateral displacement (DLD), a technique widely used in microfluidics~\cite{Huang2004, Davis2006,loutherback2009deterministic, Kim2017} and recently demonstrated to function down to nanometer scales \cite{Wunsch2016}.

\section{Computational model}\label{sec:compmod}

\subsection{Setup}\label{sec:setup}

\begin{figure}[t]
	\centering
	\begin{subfigure}[b]{.49\linewidth}
		\centering
		\includegraphics[width=.99\textwidth]{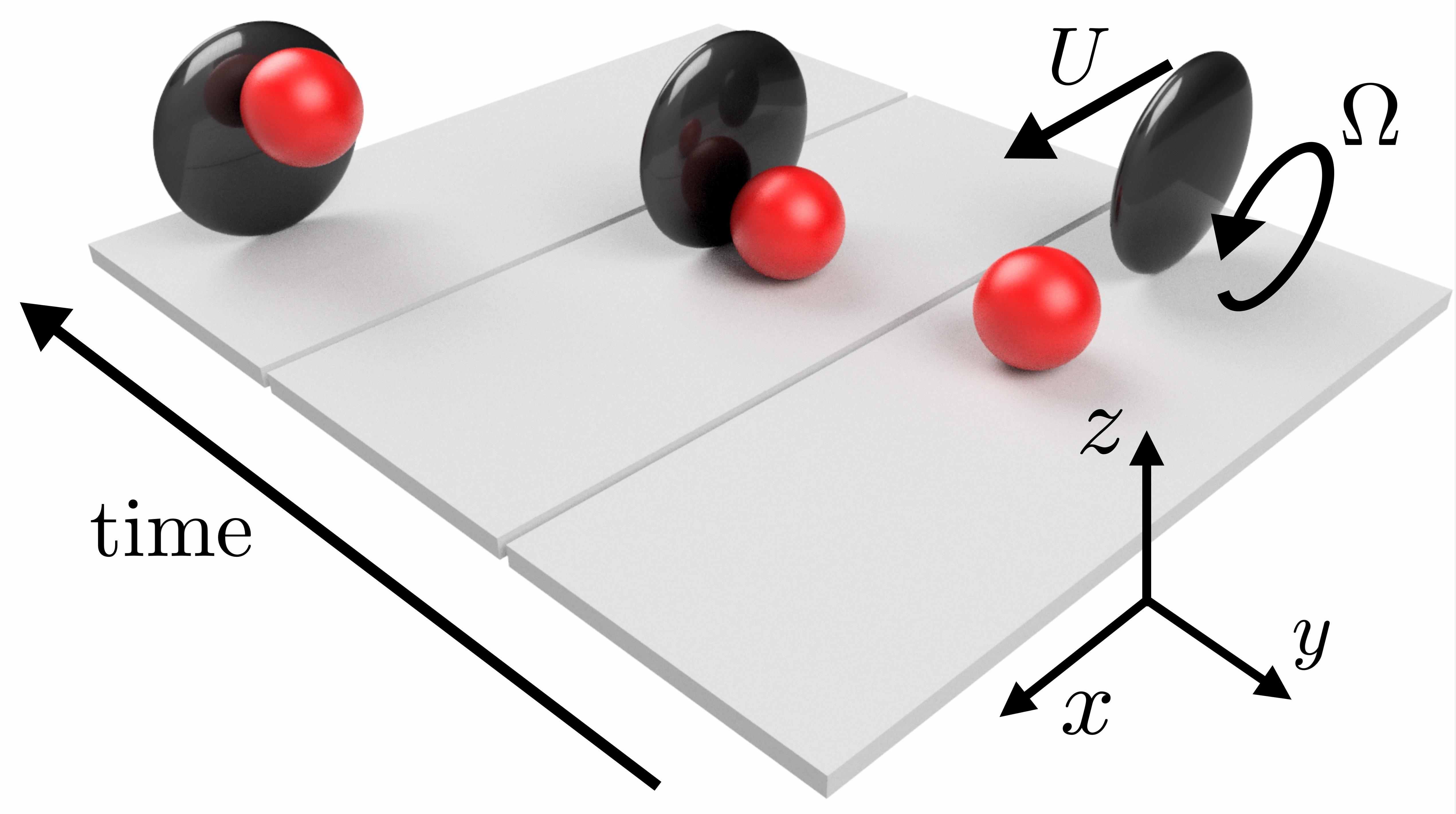}
		\caption{}
	\end{subfigure}
	\begin{subfigure}[b]{.49\linewidth}
		\centering
		\includegraphics[width=.99\textwidth]{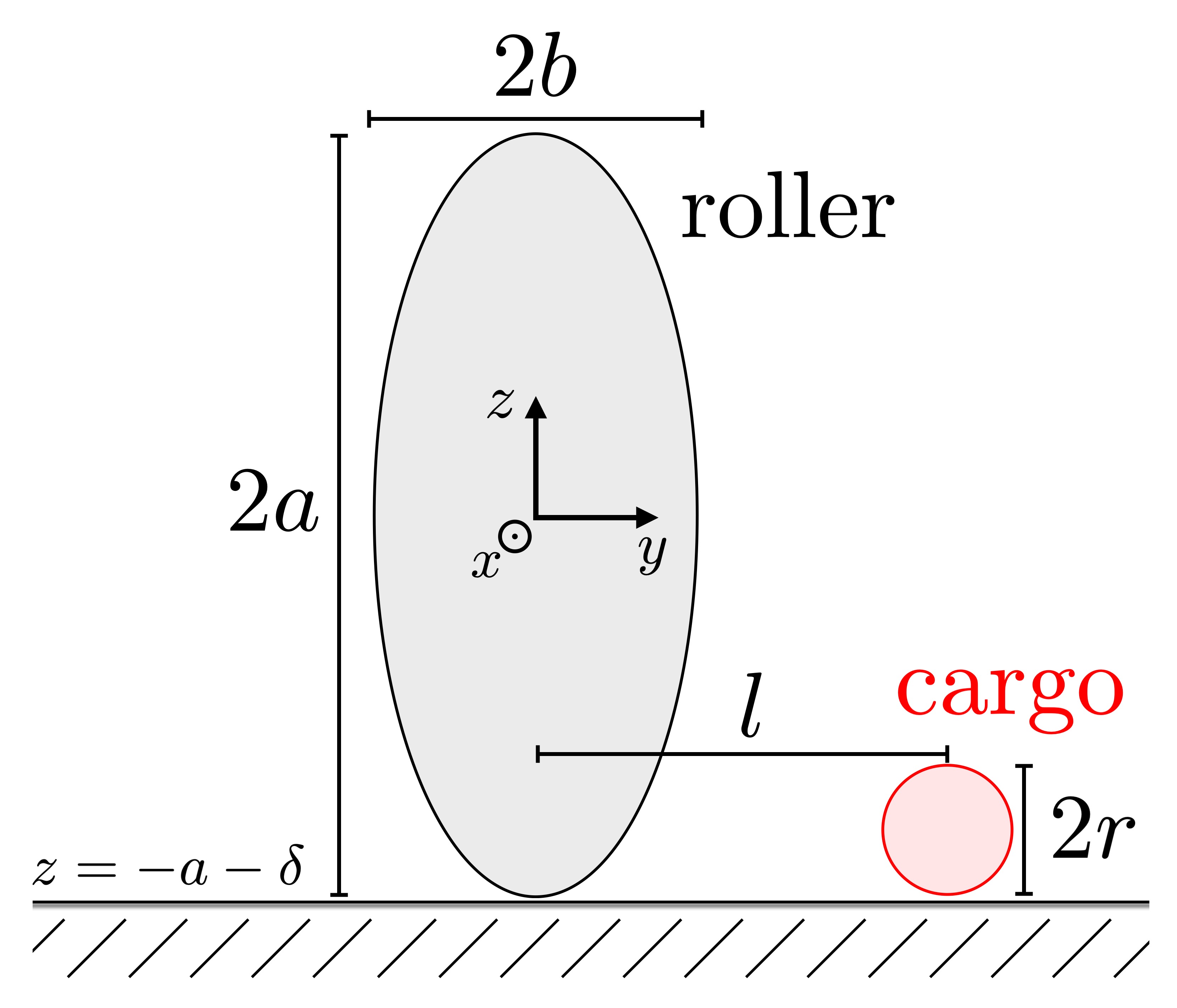}
		\caption{}
	\end{subfigure}
	\caption{(a): Schematic illustration of  {irreversible} trapping by a surface roller illustrating the trapping of a passive  {spherical} particle (cargo, red) by the rotating ellipsoidal particle (roller, black) due to steric repulsion from the bottom wall. (b): Sketch of the initial geometry projected in the $y$-$z$ plane, illustrating relevant length scales.}
	\label{fig:setup}
\end{figure}

The geometrical setup of our model problem  is illustrated in Fig.~\ref{fig:setup}. We consider Stokes flow as described by Eq.~\eqref{eq:Stokes} (i.e.~we assume the Reynolds number to be much smaller than unity) in a semi-infinite domain described by Cartesian coordinates $(x,y,z)$. The roller is modelled as an oblate spheroid with semi-major axes of length $a$ in the $x$- and $z$-directions and semi-minor axis of length $b\leq a$ in the $y$-direction, centred at $(0,0,0)$ in a frame where it is stationary (illustrated in black in the figure). A rigid boundary is placed at $z=-a-\delta$, where $\delta\ll a$ is the width of the gap between the roller and the domain boundary, which is non-zero due to the presence of a lubrication film in creeping flow. The roller translates with a velocity $\bm{U}=U\hat{\bm{x}}$, rotates at a rate $\bm{\Omega}=\Omega\hat{\bm{y}}$, is force-free and subject to a fixed external torque of the form $\bm{G}=G\hat{\bm{y}}$. The cargo particle is assumed spherical with radius $r$, force- and torque-free and initially placed far ahead of the roller with its centre at height $z=r$ and displaced sideways by a distance $l$ in the $y$-direction (illustrated in red in the figure). Note that $l$ refers to the initial value of this displacement, which changes when the roller passes the cargo. Finally, we assume that the no-slip boundary condition holds on both the wall and the roller and hence that the fluid velocity matches the  velocity of the boundary.

\subsection{The finite-element routine}\label{sec:FEroutine}
Since a full dynamic simulation of the roller and the cargo in this geometry is prohibitively expensive, we approximate the dynamics by calculating the flow due to the roller alone and determine the trajectories of the cargo under the assumption that it is sufficiently small not to create a significant disturbance to the velocity field. This approximation is exact in the case of a vanishingly small cargo particle and approximately correct for small values of  $r/a$.

In Stokes flow, the hydrodynamic force $\bm{F}$ and torque $\bm{G}$ acting on the roller are related to its translational velocity $\bm{U}$ and angular velocity $\bm{\Omega}$ by an instantaneous linear relation of the form
\begin{align}
	\begin{pmatrix}
	\bm{F} \\
	\bm{G}
	\end{pmatrix}
	=-
	\begin{pmatrix}
	\bm{A} & \bm{B} \\
	\bm{C} & \bm{D}
	\end{pmatrix}\cdot
	\begin{pmatrix}
	\bm{U} \\
	\bm{\Omega}
	\end{pmatrix},
\end{align}
where $\bm{A}$, $\bm{B}$, $\bm{C}=\bm{B}^T$ and $\bm{D}$ are positive definite matrices that depend on the instantaneous position and orientation of the spheroid. Their combination is called the resistance tensor \cite{happel2012low}. For a spheroid with the orientation described above, classical symmetry arguments allow one to deduce that the components of the resistance tensor associated with translation in the $x$-direction and rotation about the $y$-axis decouple from the others and only give rise to forces in the $x$-direction and torques in the $y$-direction (in other words, the resistance tensor is block-diagonal). In particular, there can be no motion in the $z$-direction and consequently the associated components of the resistance tensor are constant in time. 

In order to determine the values of the resistance tensor, we use a finite-element routine (COMSOL Multiphysics version~4.4) and compute the flow field $\bm{u}(\bm{x})$ due to an ellipsoid in this geometry with prescribed translational velocity and zero orientational velocity, and vice versa. In both cases, we compute the hydrodynamic force and torque on the ellipsoid according to
\begin{align}
	\bm{F}=\int\!\!\!\int \bm{\sigma}\cdot\bm{n}\,\text{d}S,\quad	\bm{G}=\int\!\!\!\int \bm{x}\times\bm{\sigma}\cdot\bm{n}\,\text{d}S,
\end{align}
where $\bm{\sigma}=-p\bm{I}+\mu\left(\nabla\bm{u}+(\nabla\bm{u})^T\right)$ is the hydrodynamic stress tensor, $\bm{n}$ the unit outward normal to the roller surface, $\bm{x}$ the position vector and the integral is taken over the surface of the ellipsoid. Exploiting linearity, this allows us to invert (the relevant part of) the resistance tensor and thus find the translational velocity $\bm{U}=U\hat{\bm{x}}$ and orientational velocity $\bm{\Omega}=\Omega\hat{\bm{y}}$ for a given value of  the applied torque, $G$, the roller aspect ratio, $b/a$, and the relative gap width, $\delta/a$.

\begin{figure}[t]
	\centering
	\includegraphics[width=0.8\linewidth]{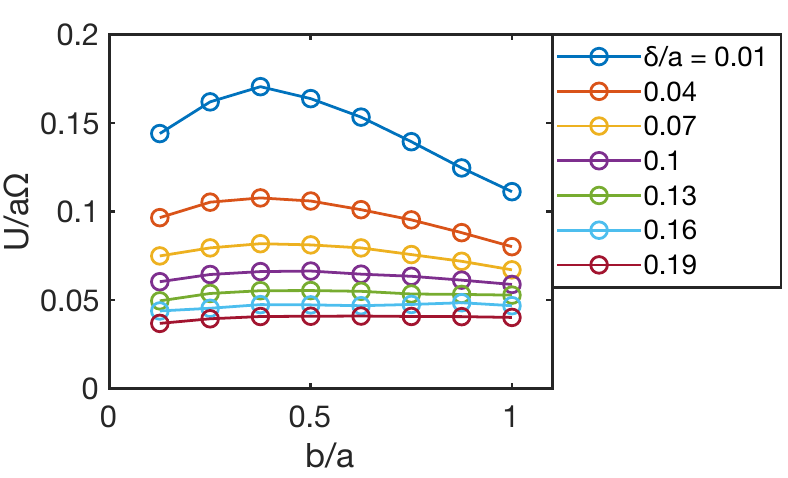}
	\caption{Computed rotation-translation coupling rate, $U/a\Omega$, for an ellipsoid subject to a constant torque, $G\hat{\bm{y}}$, as a function of its aspect ratio, $b/a$, for a range of dimensionless gap widths, $\delta/a$ (see notation in Fig.~\ref{fig:setup}).}\label{fig:couplingrate}
\end{figure}

In order to validate our code we compare the hereby obtained values for the ratio of $U$ and $\Omega$ to values in the literature derived theoretically using bipolar coordinates \cite{o1964slow,dean1963slow} and find good quantitative agreement (see \S\ref{sec:app1} for details). From dimensional analysis, we may deduce that we can write the coupling rate as $U/a\Omega= \gamma(\delta/a,b/a)$ where $\gamma$ is a dimensionless function of two dimensionless variables. This is illustrated in Fig.~\ref{fig:couplingrate}, where we see that the coupling rate depends only weakly on the precise value of the parameters unless $\delta/a$ becomes very small. This is in agreement with lubrication theory, which predicts a divergence as $\delta\to0$ \cite{goldman1967slow}. In a similar fashion, we expect the flow field and trapping dynamics to be robust  against variations of the parameter $\delta$, as long as $\delta/a\gtrsim0.04$. In what follows, we shall therefore limit our computational analysis to the case $\delta/a=0.04$.

\subsection{Simulating cargo trajectories}\label{sec:simcargotrajs}
Next we consider the trajectory of a force- and torque-free spherical cargo particle in the flow field created by the roller. Fax\'en's first law states that the velocity, $\bm{V}$, of a force-free spherical body with radius $r$ in an unbounded Stokes flow $\bm{u}(\bm{x})$ is given by \cite{kim2013microhydrodynamics}
\begin{align}
	\bm{V}=\left(1+\frac{r^2}{6}\nabla^2\right)\bm{u}(\bm{x}).
\end{align}
We note that this formula is only exact in an unbounded geometry, while near a wall that there are corrections of $\mathcal{O}(r^2/d^2)$ where $d$ is the distance between the centre of the cargo and the closest boundary. Furthermore, the relative size of the Laplacian term is $\mathcal{O}(r^2/L^2)$, where $L$ is the typical length scale of variations in the flow velocity. We make a simplifying assumption here and neglect both these terms, so that $\bm{V}=\bm{u}(\bm{x})$ and therefore approximate cargo trajectories may be obtained by integrating streamlines of the flow created by the roller. Mathematically this corresponds to the limit $r^2\ll d^2,L^2$.  

A crucial step towards modelling the lateral migration of particles is taking into account 
 the   steric interactions between the lower boundary and the roller.  With the aim to remove the velocity component normal to the boundary and thus model steric repulsion with no friction, we integrate trajectories according to
\begin{equation}\label{eq:steric}
\frac{d\bm{x}}{dt}=\begin{cases}
\left(\bm{1}-\bm{n}\bm{n}\right)\cdot\bm{u}(\bm{x}) &\text{if cargo is in contact with boundary,}\\
\bm{u}(\bm{x}) &\text{if not,}
\end{cases}
\end{equation}
where $\bm{n}(\bm{x})$ is a unit normal vector at the point of contact for any position $\bm{x}(t)$ of the cargo centre such that the cargo touches a boundary (which is permitted to be either the wall or the roller). { In appendix \S\ref{sec:app2}  we examine the accuracy of this model by comparing it with detailed finite-element simulations at judiciously chosen values of the model parameters, and demonstrate  its relevance for the modelling of our problem.}

Numerically, we initialise the cargo centre at position $(3a,l,-a+r)$ in the frame where the roller is stationary and centred at $(0,0,0)$ and solve for the cargo trajectory, $\bm{x}(t)$, for various values of the roller aspect ratio $b/a$, the relative cargo size $r/a$ and the initial lateral displacement $l/a$. To this end, we use a   forward-Euler scheme with a time-step   sufficiently small for the results to be robust to variations in step size by a factor of two. Note that since the flow is linear in $G$, the value of the applied torque has no influence on the geometry of particle trajectories and it only determines the overall magnitude of the flow field.

\section{Computational results}\label{sec:comres}

\subsection{Phase diagram for cargo entrapment}\label{sec:phasediag}

\begin{figure}[t]
	\centering
	\includegraphics[width=0.8\linewidth]{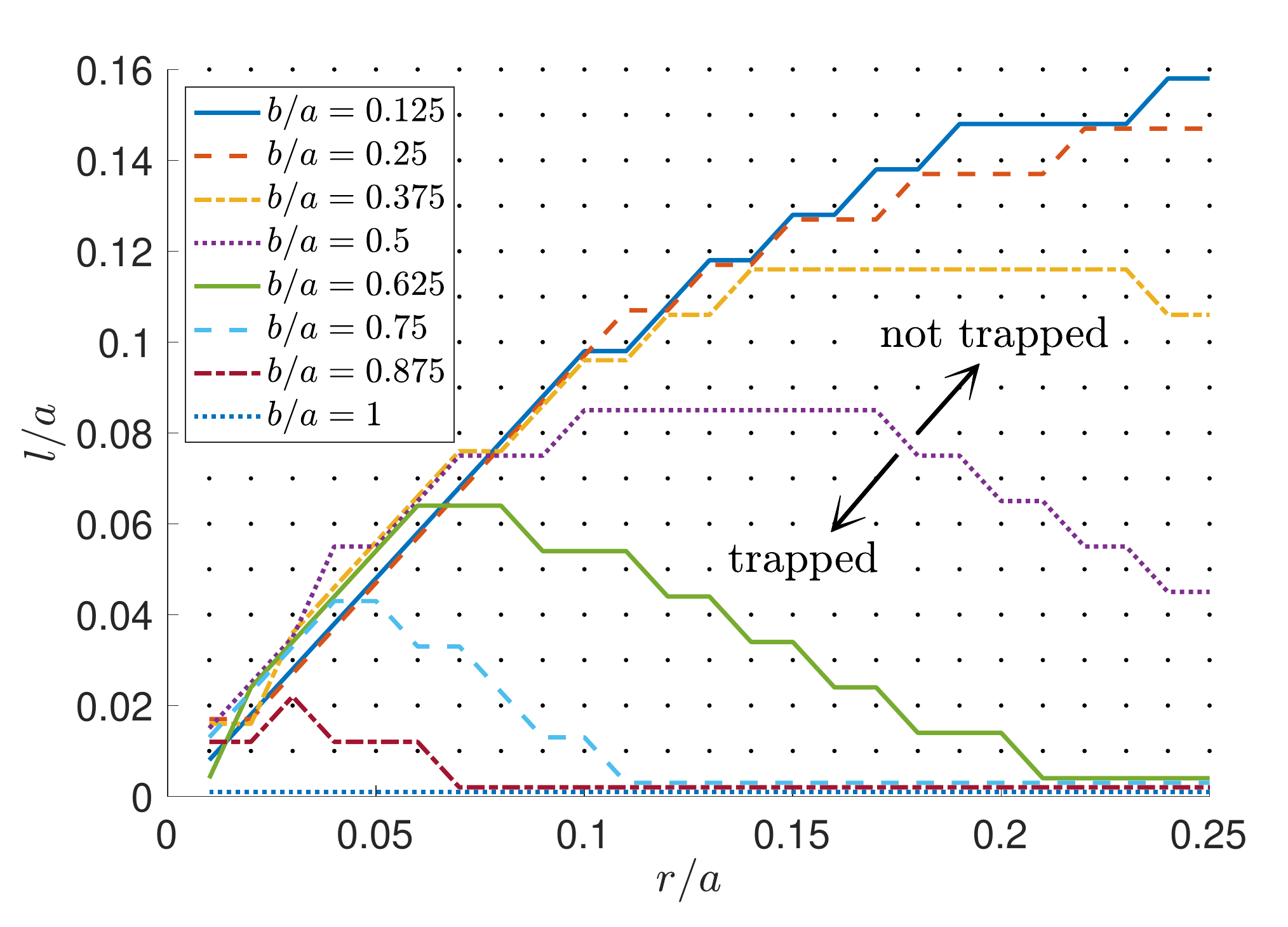}
	\caption{Phase diagram for entrapment of cargo particles by rollers with different aspect ratios, $b/a$. Each black dot corresponds to a numerically simulated parameter configuration of relative cargo size $r/a$ and initial lateral displacement $l/a$. For configurations above a given curve no trapping occurred, while for parameters below the curve the cargo particle was trapped in a periodic trajectory in the frame of the roller.}\label{fig:phasediag}
\end{figure}

Numerically, we probe the parameter space of $r/a$ between $0.01$ and $0.25$ and $l/a$ between $0.01$ and $0.30$ in increments of $0.01$ for eight different values of the aspect ratio $b/a$ between $0.125$ and $1$ in increments of $0.125$. Our  computational results  are summarised in the phase diagram shown in Fig.~\ref{fig:phasediag}. Each black dot corresponds to a single simulation and the coloured lines indicate the boundary between   configurations for which trapping of the cargo by the roller vortex is observed (below) and not observed (above). Here entrapment is defined as the convergence of the cargo to a periodic orbit in the frame where the roller is stationary. Since no entrapment occurs for $l/a\geq0.16$ we omit this range in the diagram for clarity. { Furthermore, we observe that steric interactions generally occur only between the cargo and the lower wall, but never between cargo and roller. For all values in our examined parameter range, the flow induced by the roller advects the cargo sufficiently far to the side to prevent this situation.}

We draw four main conclusions from the data summarised in Fig.~\ref{fig:phasediag}. First, and most obviously, the range of cargo sizes ($r/a$) and initial lateral displacements ($l/a$) that lead to entrapment decreases as the aspect ratio is varied from a very flat ellipsoid ($b/a=0.125$) to a sphere ($b/a=1$). In fact, in the case of a sphere   no trapping is observed at all. These results suggest that a narrow aspect ratio is conducive to trapping. Secondly, the dependence on the initial cargo position,  $l/a$, is monotonic for every configuration of the other parameters, with a well-defined threshold above which no trapping occurs. This result makes intuitive sense, since a cargo particle placed very far to the side of the roller will experience little to no deflection, while particles in the path of the roller experience the strongest flows. Thirdly, and perhaps most importantly, we observe that the dependence of the trapping threshold on cargo size $r/a$ is not monotonic. Instead, a well-defined range of values exists for each configuration of lateral placement $l/a$ and roller aspect ratio $b/a$ in which trapping occurs. Therefore this provides a constraint on what type of cargo a given roller can trap and transport at all, since only cargo of the right size will be pushed from its unbounded trajectories into a region of closed streamlines. Finally, we see in Fig.~\ref{fig:phasediag} that the slope of the separatrices for small values of $l/a$ and $r/a$ are all approximately one, regardless of the roller aspect ratio. This suggests that in order to be trapped, a cargo particles must not lie entirely on one side of the plane of symmetry of the roller.

\subsection{Illustration of  cargo migration}\label{sec:illust}
\begin{figure}[t]
	\centering
	\begin{subfigure}[b]{.4\linewidth}
		\centering
		\includegraphics[width=.99\textwidth]{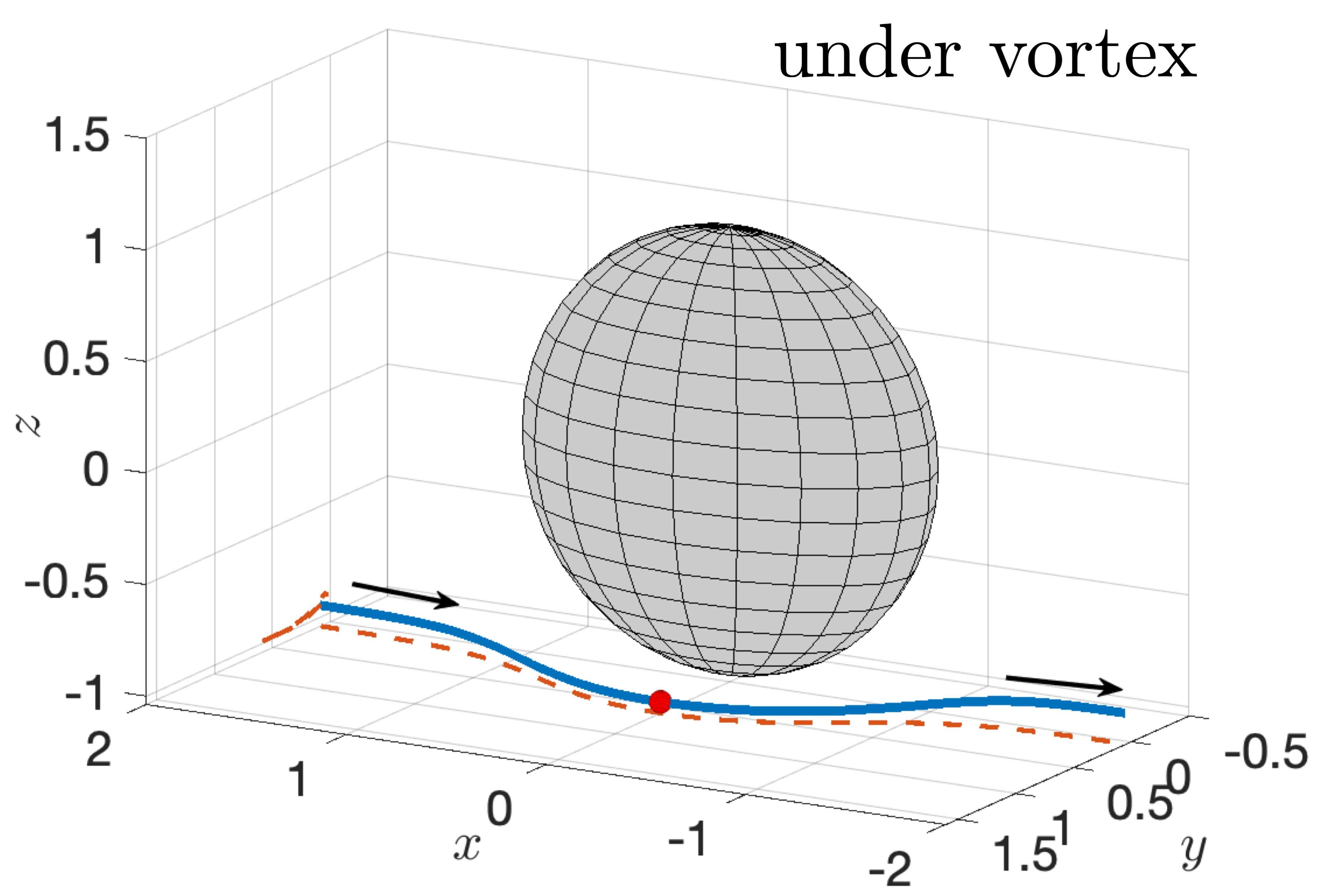}
		\caption{$r/a=0.05$}
	\end{subfigure}
	\begin{subfigure}[b]{.4\linewidth}
		\centering
		\includegraphics[width=.99\textwidth]{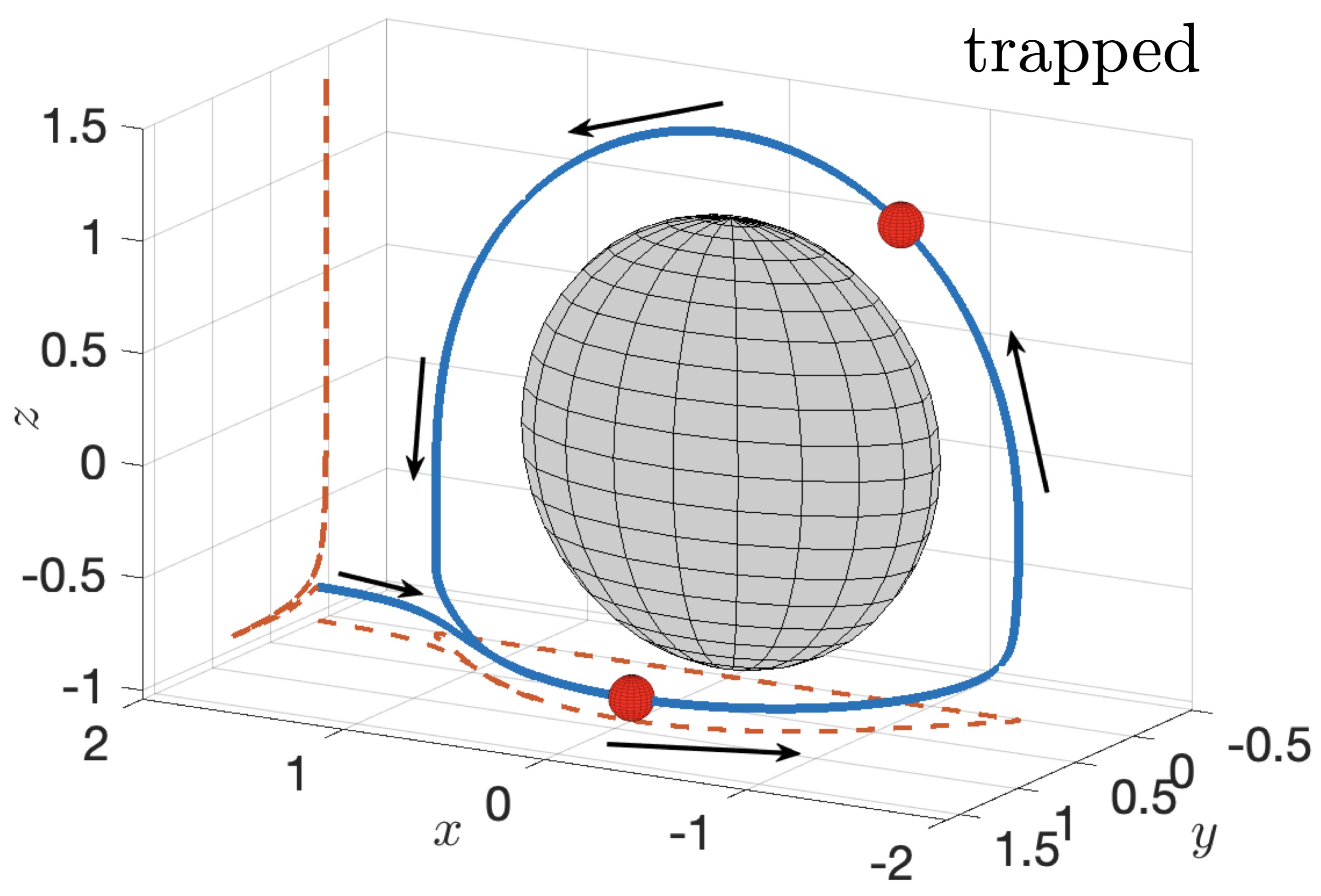}
		\caption{$r/a=0.10$}
	\end{subfigure}
	\begin{subfigure}[b]{.4\linewidth}
		\centering
		\includegraphics[width=.99\textwidth]{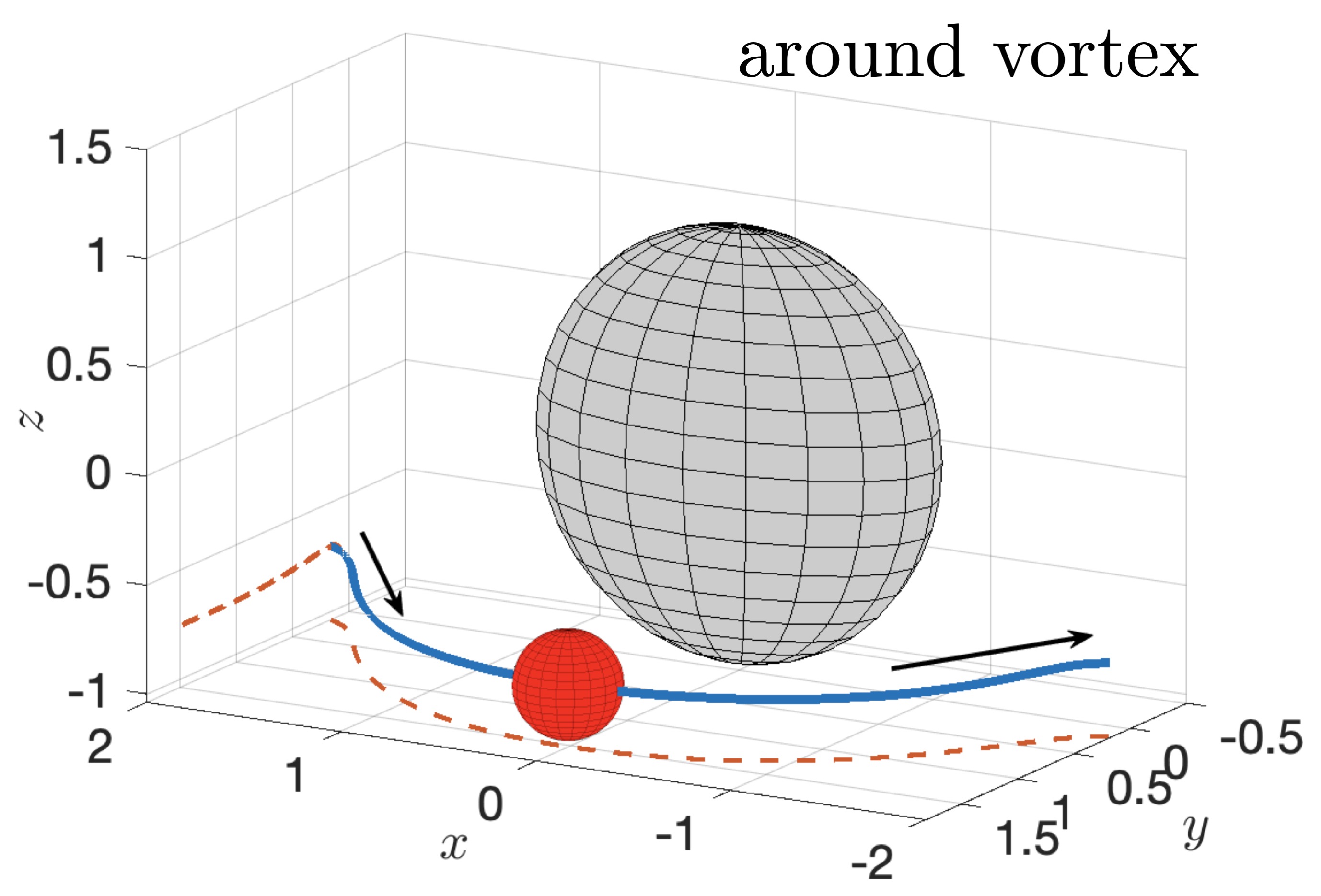}
		\caption{$r/a=0.25$}
	\end{subfigure}
	\begin{subfigure}[b]{.4\linewidth}
		\centering
		\includegraphics[width=.99\textwidth]{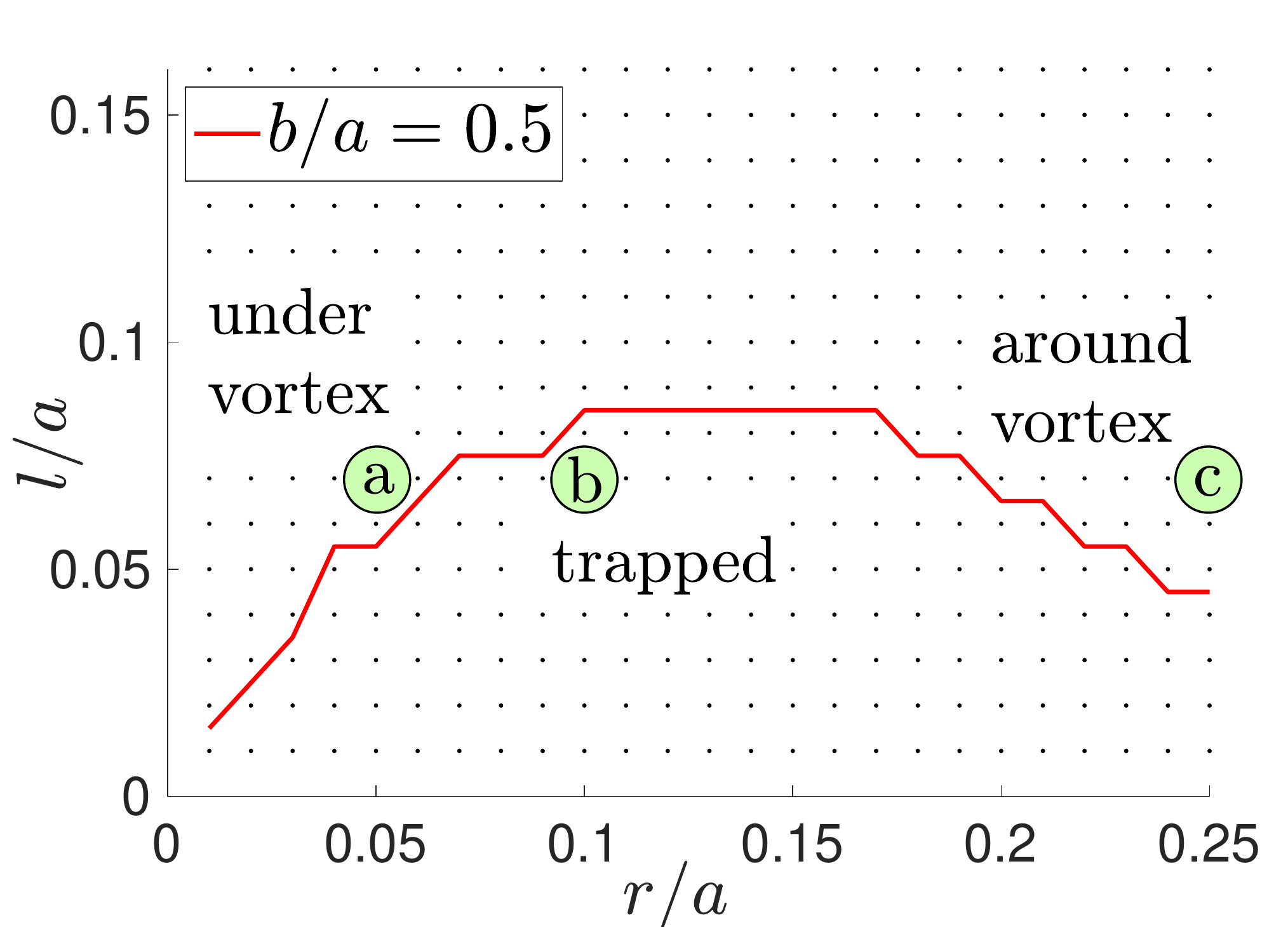}
		\caption{}
	\end{subfigure}
	\caption{Numerical illustration of trapping for a roller (grey) of aspect ratio $b/a=0.5$ and spherical cargo (red) of initial lateral displacement $l/a=0.07$ and of three different sizes $r/a$. The thick blue line show the cargo trajectory in the frame where the roller is stationary, with the dashed red lines indicating shadows on planes perpendicular to the $x$- and $z$-axes added for clarity. The roller travels in the positive $x$-direction. Arrows indicate the direction of the cargo trajectories and axes are scaled by $a$.
	(a): Small cargo particles are squeezed through under a region of closed streamlines;
	(b): Medium-sized particles are pushed into a vortex of closed streamlines through steric interactions with the bottom wall and are therefore trapped;
	(c):  Large particles are confined to unbounded trajectories around the vortex;
	(d): Location of (a)-(c) in the phase diagram in Fig.~\ref{fig:phasediag}.}
	\label{fig:illustration}
\end{figure}

In order to shed  more light on the entrapment mechanism, we illustrate in Fig.~\ref{fig:illustration} three exemplary parameter configurations. We choose the values $b/a=0.5$ and $l/a=0.07$, which can be seen in Fig.~\ref{fig:phasediag} (and as reproduced in Fig.~\ref{fig:illustration}d) to feature different behaviour for three different ranges of $r/a$, for which we select the values $r/a=0.05$, $0.10$ and $0.25$. An examination of the trajectories reveals that a cargo particle that is too small squeezes through below a region of closed streamlines next to the roller (Fig.~\ref{fig:illustration}a), while cargo that is too large is instead lifted up to trajectories around the same region (Fig.~\ref{fig:illustration}c). For a particle of intermediate size (Fig.~\ref{fig:illustration}b), we observe a trajectory that is not fore-aft symmetric, as the cargo is pushed into a vortex of closed streamlines by steric interactions with the bottom wall. 
   
\subsection{Pure translation does not lead to trapping}\label{sec:transnotrap}
In a similar fashion to the case of a roller that is subject to a constant torque, we also examined the case of a purely translating spheroid with $\Omega=0$ and $U\neq 0$. Such a scenario is somewhat artificial, since it requires a finely tuned ratio of non-zero force and torque, but is nonetheless instructive to examine because it exhibits strikingly different behaviour.  We considered the cases of a sphere ($b/a=1$) and a very flat spheroid ($b/a=0.125$) for the range $l/a=0.01-0.30$ and $r/a=0.01-0.25$ in steps of $0.01$ each and integrate streamlines numerically in the same fashion as above. In this case, we find that for no parameter value in this range the cargo particle is trapped, instead it always passes the spheroid on a nearly unperturbed trajectory. From this we can hence conclude that the rotation of the spheroid is essential for  entrapment and bounded transport of cargo particles.

\section{Theoretical model}\label{sec:theory}

The phase diagram  in Fig.~\ref{fig:phasediag} was obtained by simulating the trajectories of cargo particles numerically. Since the geometry of the problem is rather complex, we propose two different theoretical models that each focus on a different feature of the numerical observations. First, we consider the flow induced by a rotating and translating disc in an unbounded fluid (neglecting the influence of the wall), in order to explain why trapping is more pronounced for flat rollers and why there is an upper limit to the size of cargo that may be trapped. We then propose a two-dimensional singularity model, to explain the physical mechanism of trapping and why no trapping is observed for a purely translating spheroid.

\subsection{Vortex flow surrounding a translating and rotating rigid disc}\label{sec:disc}

The first important feature of the phase diagram is the prominence of trapping for rollers with a narrow aspect ratio. In order to elucidate this further, we begin by considering the extreme case of a rolling disc, i.e.~we consider the limit $b=0$, and in order to make analytical progress, we ignore the presence of the wall. We consider a frame in which the disc is stationary but rotating with angular velocity $\bm{\Omega}=\Omega\hat{\bm{y}}$, and scale lengths by the disc radius, $a$. Since the no-slip condition is applied on the disc's surface, very near to it (that is for $|y|$ small) the fluid is approximately in solid body rotation. In terms of cylindrical polar coordinates $(\rho,\theta,y)$ with $\rho^2=x^2+z^2$ and $\tan\theta=x/z$ we show in appendix \S\ref{sec:app3} that the streamfunction for a rotating rigid disc in a quiescent infinite fluid is of the form $\bm{\psi}=\psi(\rho;y)\bm{\hat{y}}$ where
\begin{align}
\psi&=\frac{\Omega}{\pi}\left[-3\frac{y^2}{\lambda}+\lambda+\left(\frac{y^2}{\lambda^2}+1+3 y^2- \lambda^2\right) \cot ^{-1}\lambda\right],
\end{align}	
and
\begin{equation}
\lambda=\left\{{\frac{1}{2} \left(\rho^2+y^2-1\right)+\frac{1}{2} \left[{\left(\rho^2+y^2-1\right)^2+4
		y^2}\right]^{1/2}}\right\}^{1/2}.
\end{equation}
To model our simulations, we still need to add translation in the plane perpendicular to the axis of rotation. To this end we define the non-dimensional coupling rate between translation and rotation as $\gamma=U/a\Omega$. As illustrated in Fig.~\ref{fig:couplingrate}, the translation is slow and typically $\gamma\approx 0.1\ll 1$. In order to describe the flow topology and identify regions of closed streamlines, we would like to use the streamfunction formalism for the more complicated problem of coupled translation and rotation as well, since it is easy to identify the value of $\psi$ at stationary points, where $\nabla\psi=\bm{0}$, and then trace the contours that separate topologically distinct regions of the flow. However, as is evident from the numerics also, there is a new out-of-plane component of the flow in the $y$-direction as soon as translation is considered.

In order to circumvent this issue, we exploit the fact that $\gamma$ is small and simply add a background flow of magnitude $-U\hat{\bm{x}}$. This effectively amounts to neglecting the correction to the streamfunction due to the no-slip condition on the disc surface. It is easily seen that the magnitude of the discrepancy on the boundary is uniformly equal to $\gamma$ and thus, by linearity of Stokes flow, the global error in $\bm{u}$ incurred is also linear in $\gamma$.

After rescaling and removing an apparent divergence at $\lambda=0$ by substituting $\rho$ for $y$ we then find that the approximated translation-rotation streamfunction is hence given by
\begin{align}
\psi(x,z;y,\gamma)=\frac{1}{\pi}\left(\frac{3\lambda}{1+\lambda^2}\rho^2-2\lambda+\left(2+2\lambda^2-\frac{1+3\lambda^2}{1+\lambda^2}\rho^2\right) \cot ^{-1}\lambda\right)+\gamma z.
\end{align}

\begin{figure}[t]
	\centering
	\begin{subfigure}[b]{.49\linewidth}
		\centering
		\includegraphics[width=.99\textwidth]{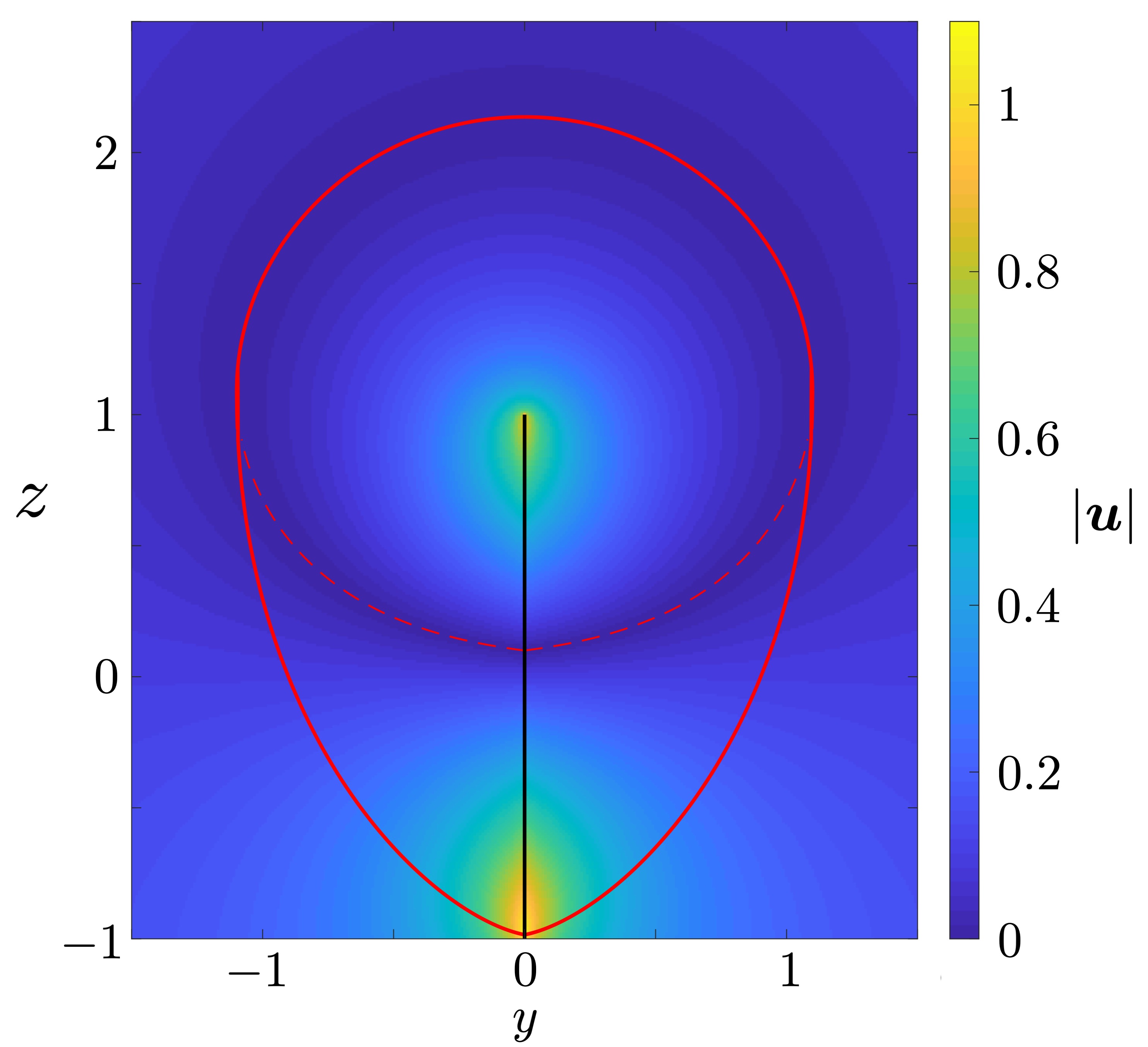}
		\caption{$y$-$z$ plane}\label{fig:platevortexa}
	\end{subfigure}
	\begin{subfigure}[b]{.49\linewidth}
		\centering
		\includegraphics[width=.99\textwidth]{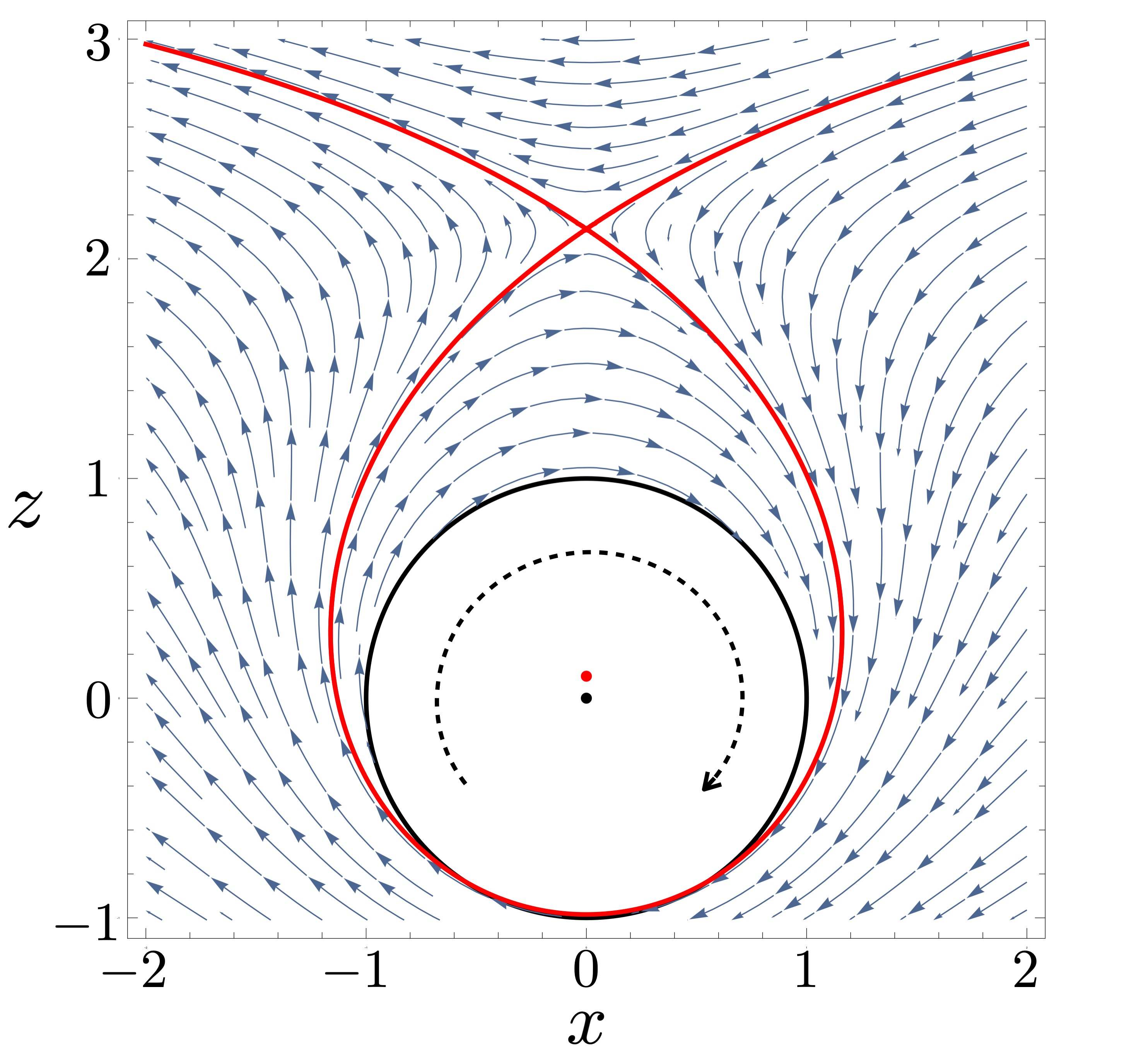}
		\caption{$x$-$z$ plane}\label{fig:platevortexb}
	\end{subfigure}
	\caption{Illustration of the vortex surrounding a rotating and translating disc for $\gamma=0.1$ with lengths scaled by the disc radius $a$. (a) Front view ($y$-$z$ plane):  The thick red line indicates the vortex boundary, with saddle points constituting the top half, while the dashed red line indicates two lines of centre stagnation points. (b) Side view ($x$-$z$ plane): The separatrix streamline is indicated in red, as is the centre stagnation point. The geometrical centre of the roller is indicated in black for comparison. The discrepancy between the points gives an indication of the magnitude of the error in this model.}\label{fig:platevortex}
\end{figure}

This streamfunction now allows us to identify a region of closed streamlines semi-analytically. We find numerically that for small values of $|y|$ there exist two stagnation points which are located at $x=0$ and $z$ positive. As $|y|$ increases, these vanish through a saddle-node bifurcation. By identifying contours of $\psi$ equal to the value at the saddle, we can then determine the size and shape of the vortex. For the representative value of $\gamma=0.1$ this is illustrated in Fig.~\ref{fig:platevortex} by means of two cross-sections in the planes $x=0$ and $y=0$. We see two topologically distinct regions, separated by the thick red line that corresponds to intersection of the streamlines that pass through the line of saddle points with the plane $x=0$  (Fig.~\ref{fig:platevortexa}). Inside this region, streamlines are closed and encircle the lines of centre points (dashed line), while outside the streamlines are unbounded and extend to infinity in the $x$-direction (Fig.~\ref{fig:platevortexb}). The discrepancy between the centre of the disc and the centre stagnation point of the flow is due to the approximation we made earlier and is equal to $\gamma$.

While it is expected that the presence of the wall will also alter the general topology of the flow field, the   model   illustrates that the flat geometry leads to the fluid in the region $\{\rho<1,|y|\ll1\}$ moving in nearly solid-body rotation. As seen in Fig.~\ref{fig:illustration}, this still holds true for the vortex in the presence of the wall, except very close to the boundary, where this is a small distortion in the $y$-direction. This provides a constraint on the size of cargo particles that may be trapped at all, since cargo particles exceeding the size of the vortex width cannot be trapped in it by volume exclusion. Likewise, a particle placed too far to the side of the roller will simply circumvent the vortex and not get trapped either. This explains why no trapping is observed in the phase diagram 
in Fig.~\ref{fig:phasediag} for large values of $l/a$ and $r/a$.

When the aspect ratio instead approximates that of a sphere, volume exclusion is more significant. Furthermore, due to the increased curvature of the ellipsoid, the effective solid body rotation is also less pronounced. Both of these factors contribute to the observation that trapping is less pronounced for near-spherical rollers. 

\subsection{Physical mechanism of trapping by squeezing of streamlines}\label{sec:squeezing}

\begin{figure}
	\centering
	\includegraphics[width=0.6\columnwidth]{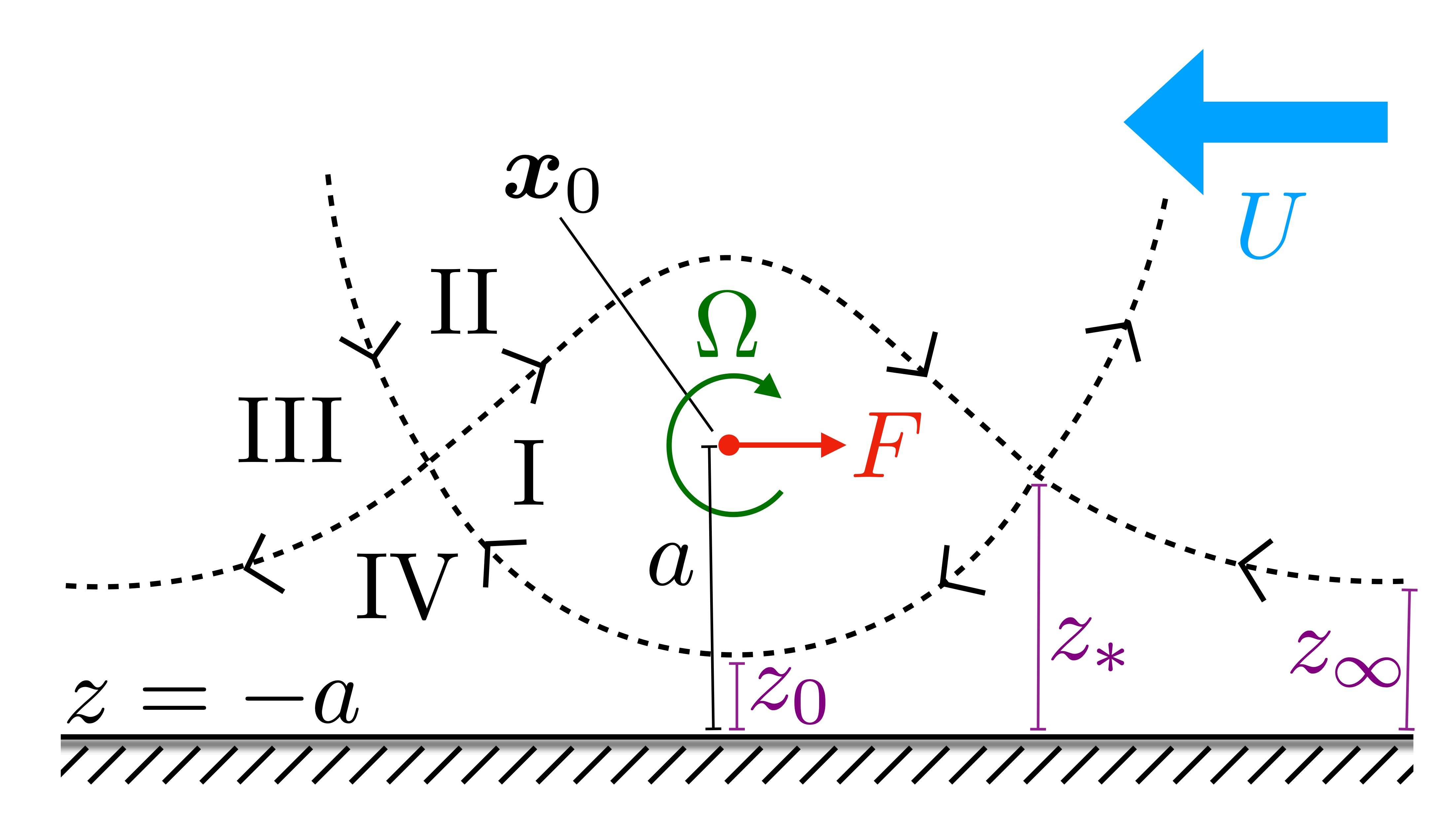}
	\caption{Sketch of the 2D model geometry. Dashed black lines correspond to separatrix streamlines dividing the flow into four topologically distinct regions I-IV described in the main text. For a force-free roller model, $F=0$.}\label{fig:sketch}
\end{figure}

As seen in Fig.~\ref{fig:illustration}, the fluid on either side of the roller is nearly in solid body rotation except very close to the boundary, where this is a distortion in the $y$-direction. In order to derive a model for the trapping mechanism we can exploit this nearly two-dimensional nature of the flow to find a 2D streamfunction, $\psi(x,z)$ whose contours approximate the flow field close to the side of the roller. Fundamentally, the flow is composed of two different components, namely  one due to the roller rotation and one due to the translation. We choose to model the rotation by means of a point (line) vortex of strength $\Omega\hat{\bm{y}}$ placed at $\bm{x}_0=\bm{0}$ above a rigid, flat no-slip surface that we place at $z=-a$. In order to model translation we add a background flow of magnitude $-U\hat{\bm{x}}$ and keep the ratio $\gamma=U/a\Omega$ as a parameter. For some additional generality we furthermore include a force per unit length of strength $F\hat{\bm{x}}$, which we set to zero for the case of a force-free roller. A sketch of the setup is shown in Fig.~\ref{fig:sketch}.

The flow due to a point line vortex next to a rigid wall differs from that of a line vortex in infinite space due to a correction that is necessary to satisfy the no-slip condition on the boundary. The method of images provides a way to interpret this correction as equivalent to the influence of  image singularities located in a hypothetical fluid on the other side of the boundary. In the case of a line vortex, these are a line vortex, a symmetric force dipole and a source-doublet placed at the mirror image point of the singularity \cite{blake1974fundamental}, just as in the case of a three-dimensional rotlet singularity. Similarly, the image of the force is given by another force, a symmetric force dipole and a source dipole. We define the streamfunction, $\psi$, such that ${\bf u}=(\partial \psi/\partial z,- \partial\psi / \partial x)$ and streamlines are lines of constant $\psi$. As described in detail in  appendix \S\ref{sec:app4} , the streamfunction is given by
\begin{equation}
\psi=\left(\eta z-1\right)\log\frac{R}{r}+\frac{2(1+\eta)(z+1)(z+2)}{R^2}-\gamma (z+1),
\end{equation}
where lengths have been scaled with $a$, $r^2=z^2+x^2$, $R^2=(z+2)^2+x^2$ and the two dimensionless parameters $\gamma$ and $\eta$ are defined as
\begin{equation}
	\gamma=\frac{U}{\Omega a},\quad \eta=\frac{F}{8\pi\mu\Omega a}.
\end{equation}
A force-free roller then corresponds to the case $\eta=0$.
As is illustrated in Fig.~\ref{fig:sketch}, the flow is divided into four topologically distinct regions for non-zero values of $\gamma$, namely (I) a vortex of closed streamlines around the singularities, (II) a counter-rotating vortex vertically above the singularities, (III) streamlines passing around the roller above and (IV) streamlines passing below. Streamlines in regions (I) and (II) are closed, while streamlines in (III) and (IV) are unbounded. The origin of these regions may be understood in terms of the actual three-dimensional geometry around the roller, in which the stagnation points in the centres of regions (I) and (II) are linked up by a vortex ring in the $y$-$z$ plane while regions (III) and (IV) are linked by streamlines circumventing the roller by bending out of the $x$-$z$ plane.

The four regions are divided by a single separatrix streamline $\psi=\psi_0$ with two stagnation points fore and aft to the roller. In the degenerate case $\gamma =0$ (no translation) these stagnation points collapse onto the wall. If $\eta=0$ and $\gamma\geq3/8$ they coalesce in a pitchfork bifurcation into a single saddle point vertically above the singularity and region (II) disappears.

For non-zero $\eta$, the position of these stagnation points is the solution to a transcendental equation. However, by means of a Taylor expansion it may be shown that for small $\gamma$ their height $z_*$ above the surface and the value of $\psi_0$ are given by
\begin{equation}
	z_*=\frac{2(1+\eta)}{(1+2\eta)^2}\gamma+\mathcal{O}(\gamma^2),\,\quad \psi_0=-\frac{1+\eta}{(1+2\eta)^2}\gamma^2+\mathcal{O}(\gamma^3).
\end{equation}
With this, we can trace the height of the separatrix to its value centrally below the roller $z_0$ and far away $z_\infty$, which are found to be
\begin{equation}
\label{eq:z_inf}
z_0=\frac{1}{2+4\eta}\gamma+\mathcal{O}(\gamma^2),\,\, z_\infty=\frac{1+\eta}{(1+2\eta)^2}\gamma+\mathcal{O}(\gamma^2).
\end{equation}
We see that their ratio obeys
\begin{equation}
\frac{z_0}{z_\infty}=\frac{1+2\eta}{2+2\eta}\leq 1.
\end{equation}
Therefore we have shown theoretically that  the separatrix streamline is squeezed for any finite value of $\eta$, i.e.~any flow with a rotational component regardless of any forcing. As $\eta\to\infty$ and there is only a force and no rotation, no squeezing of the streamlines occurs. 

In order to understand the consequences of this consider a cargo particle of radius $r\lesssim z_\infty$ resting in the path of the roller at a height less than $z_\infty$ from the wall. The particle will be advected by the flow, first towards the stagnation point, and then below the roller. If $r\lesssim z_0$ the particle will survive the squeezing and will escape on the other side of the roller, to be advected away. In contrast, if $r\gtrsim z_0$ the steric repulsion between the cargo particle and the wall means that the cargo will not survive the squeezing. Instead, it will experience a time-irreversible migration across the separatrix streamline into the vortex surrounding the flow singularities. Since the streamlines in this vortex are closed, such a particle will then remain trapped forever thereafter. This is the physical mechanism for trapping of cargo particles.

\begin{figure}[t]
	\centering
	\begin{subfigure}[b]{.49\linewidth}
		\centering
		\includegraphics[width=\textwidth]{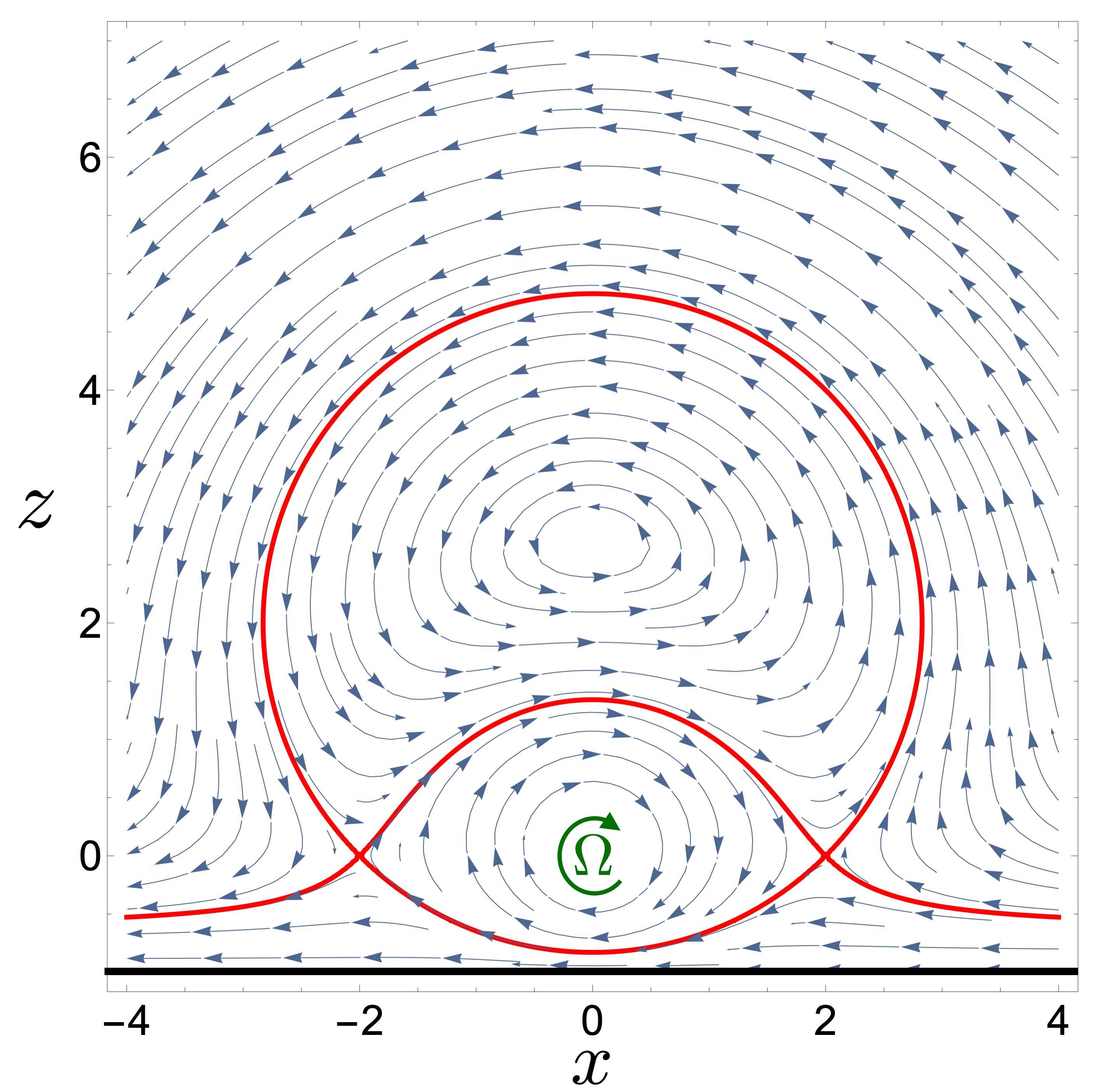}
		\caption{$\eta=0$, $\gamma=0.25$}\label{fig:wf}
	\end{subfigure}
	\begin{subfigure}[b]{.49\linewidth}
		\centering
		\includegraphics[width=\textwidth]{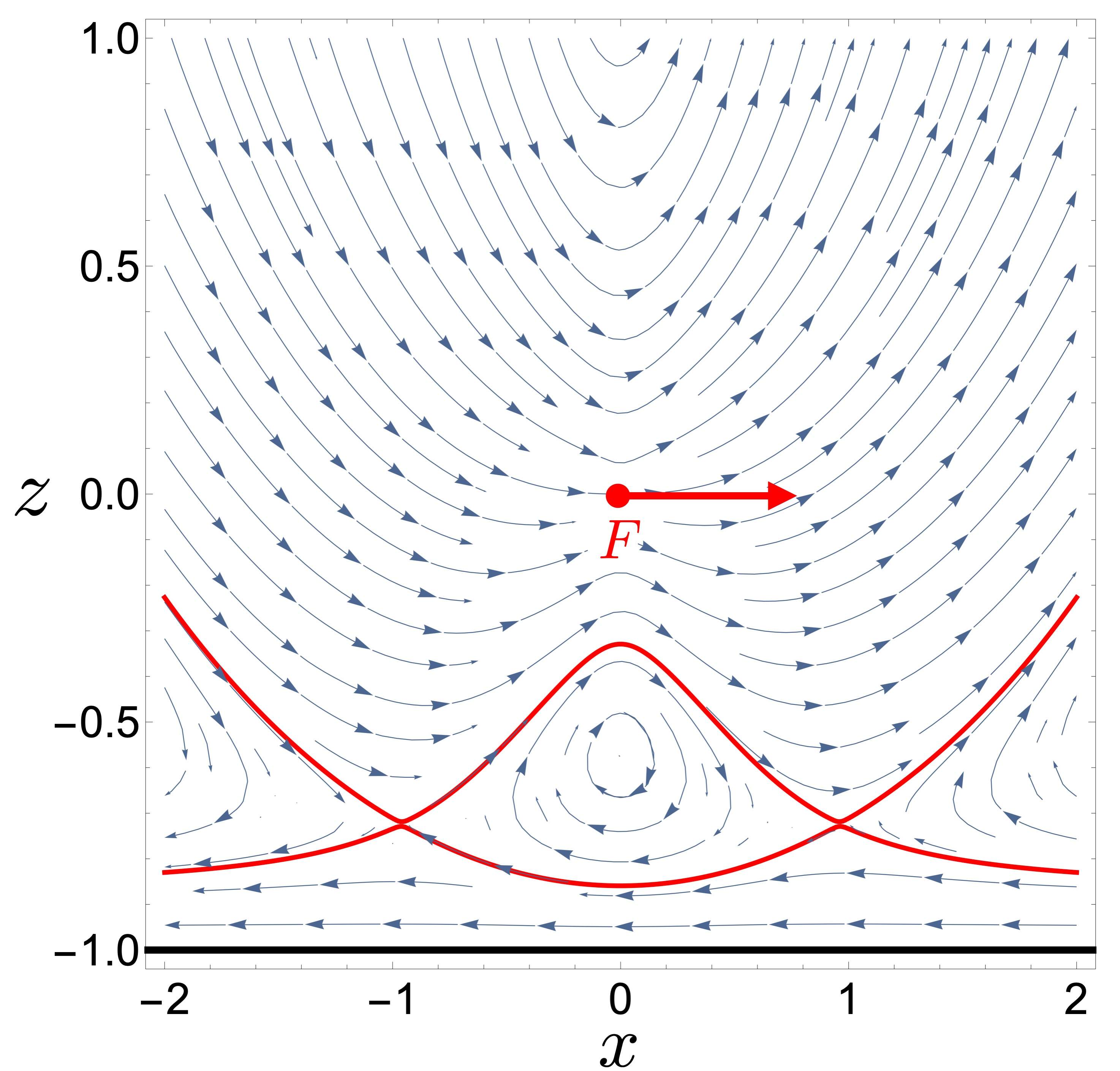}
		\caption{$\psi/\eta$ for $\eta\to\infty$ and $\gamma/\eta=0.5$}\label{fig:of}
	\end{subfigure}
	\caption{Illustration of the squeezing of streamlines in the two-dimensional   singularity model, with parameter values chosen to emphasise important features of the flow field.
	(a): In the presence of a rotation, the separatrix streamline (bold, red) is squeezed below the singularity.
	(b): For a translating force with no rotation there is still a region of closed streamlines but no squeezing occurs.}\label{fig:2dstreamlines}
\end{figure}

We note that $z_0/z_\infty$ is minimised for $\eta=0$, that is a force-free roller. Furthermore, since the squeezing requires $z_0/z_\infty${$< 1$}, it is not sufficient to have $\eta=\infty$, i.e.~pure translation. Therefore rotation of the roller is a necessary ingredient for squeezing, even though it is not for the flow topology (see illustration in Fig.~\ref{fig:of}). This agrees with our numerical observations, where no trapping occurs for a purely translating ellipsoid and any choice of parameters.

\section{Discussion}\label{sec:discussion}
In this paper we showed that the onset of hydrodynamic trapping by a surface roller near a wall is due to the physical contact of a passive finite-sized particle to the bottom wall, which breaks the time reversibility of the system. The flow field around a rotating and translating rigid body features a vortex of closed streamlines, in which particles can be trapped. However, the migration from unbounded streamlines into the vortex in the absence of gravity is only possible due to steric repulsion. While the actual three-dimensional flow field is rather complex, a simple two-dimensional singularity model allowed us to explain why rotation is the essential factor that contributes to trapping. A passive particle with a radius larger than the height of the deflected streamline experiences a steric repulsion from the bottom wall and translates into a vortex.

\begin{figure}[h]
		\centering
		\includegraphics[width=.5\textwidth]{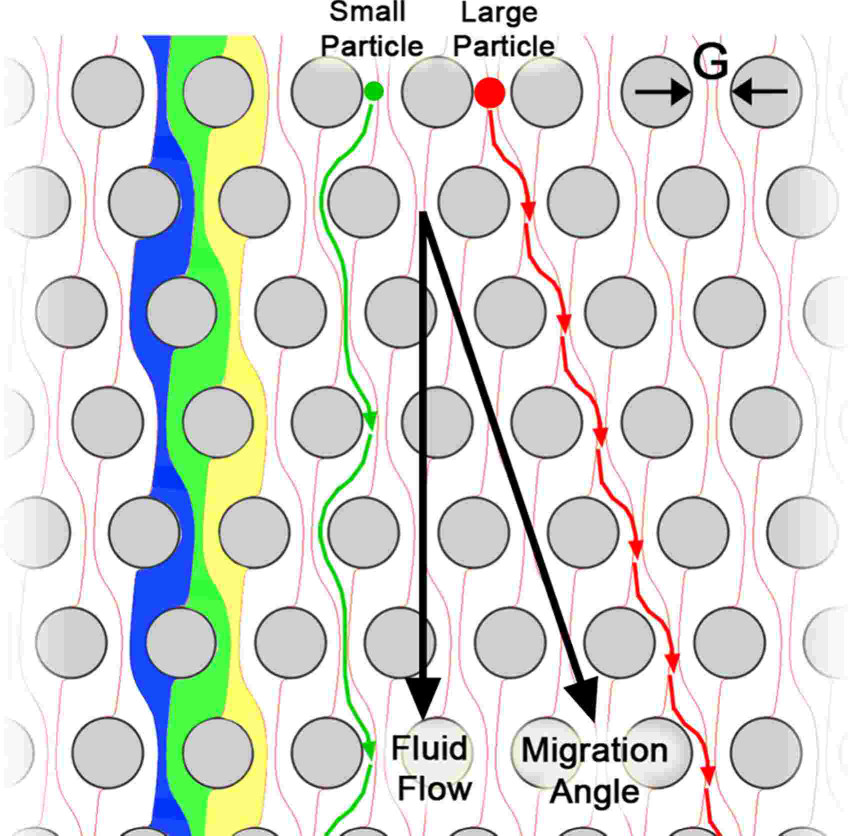}
		\caption{Schematic illustration of deterministic lateral displacement (DLD).
			While small particles (green) approximately follow the flow streamlines, large particles (red) divert their trajectories due to steric interactions with a micro-pillar array. Reprinted from Ref.~\cite{inglis2009efficient}, with the permission of AIP Publishing.}\label{fig:DLD}
\end{figure}

The trapping mechanism  proposed here is analogous to  the deterministic lateral displacement commonly used in micro-and nanofluidic separation systems~\cite{Huang2004, Davis2006, loutherback2009deterministic, Wunsch2016,Kim2017}. In a typical deterministic lateral displacement (DLD) device, particle trajectories are controlled by the steric interaction between particles and pillar arrays, as shown schematically in Fig.~\ref{fig:DLD}. Here small particles are approximately advected with the flow, but large particles collide with the array and migrate across streamlines. As a consequence, differently sized particles can be sorted by their size. Similarly, in our model only particles that are sufficiently large to experience steric interactions with the bottom wall but small enough to fit into the vortex can be selectively trapped inside of it.

In our approach to modelling this problem we made a few important assumptions. First, we assumed for computational feasibility that a finite-sized particle follows the streamlines of an isolated roller near a wall. In reality there are corrections due to the distortion of the flow field in the presence of the particle, and by Fax\'en's law also the fact that finite-sized particles do not exactly follow streamlines. { Detailed numerical calculations at judiciously chosen values of the parameters are reproduced in the appendix \S\ref{sec:app2}. They   support 
the accuracy of our model  and show that it is able to capture the key feature that enables DLD-like trapping in the entire parameter range considered in this paper. Nevertheless, due to lubrication forces the velocity of approach to the boundary is modified, and a more detailed analysis might be necessary to obtain certainty for a particular parameter configuration. In an experimental system, further deviations may be induced by the effects of thermal noise when the roller and cargo particles are sufficiently small.}

Secondly, in order to remain analytically tractable our minimal theoretical models contain many simplifications of the real problem. Indeed, our numerics show that the three-dimensional nature of the geometry and the no-slip condition on the roller surface generate a velocity field that is more complex than either of the theoretical models predict. However, for a range of parameters the essential aspects flow topology are revealed to be similar to the 2D case, with regions containing closed vortical structures present at the sides of the roller and squeezed streamlines beneath. 

Finally, we assumed that the passive particles are neutrally buoyant to eliminate the effect of gravity for simplification. In a typical experiment, the trapped objects are polystyrene particles or biological cells, which are slightly heavier than the surrounding fluid (water). In this case, the sedimentation of passive particles can be another irreversible force and induce trapping. However, this gravity-induced time-irreversibility becomes less significant in the case of a rapidly rotating roller, while the squeezing of streamlines and the thereby induced lateral migration is always present.

\appendix

\section{Verification of the finite-element method}\label{sec:app1}
\begin{figure}[h]
	\centering
	\includegraphics{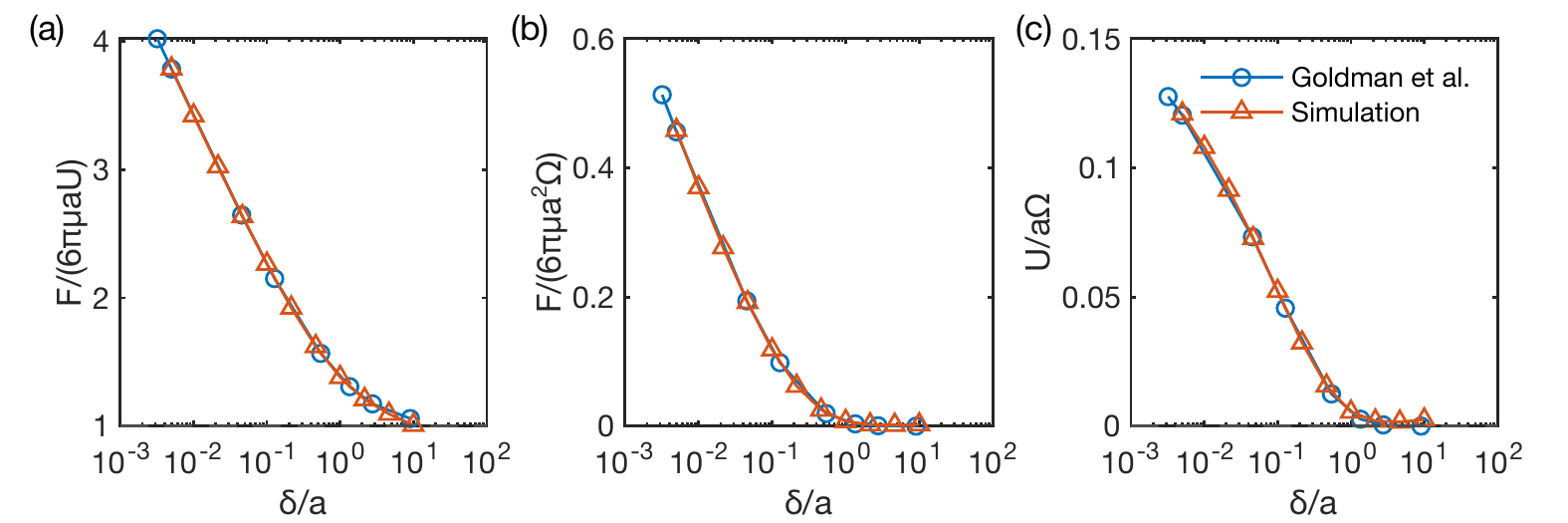}
	\caption{Comparison of our numerical method with the results by Goldman et al. for $a/b =1$ \cite{Goldman1967}.
		(a) Normalised force on a translating sphere, (b) normalised force on a rotating sphere, and (c) force-free rotation-translation coupling rate as a function of the dimensionless gap width $\delta/a$.}\label{fig:comparison_goldman}
\end{figure}

In order to verify the numerical accuracy of our finite-element routine, we simulated the force on the translating and rotating ellipsoids, calculated the rotation-translation coupling rate, and compared these results with the data obtained numerically by Goldman \textit{et al.} \cite{Goldman1967} in Fig.~\ref{fig:comparison_goldman}, which agrees with theoretical predictions obtained using bispherical coordinates \cite{o1964slow,dean1963slow}.
The mesh was refined until the deviation from the Goldman's data fell below 1\% at the gap width $\delta/a=0.005004$.

{
	
	\section{Analysis of the flow disturbance due to finite cargo size}\label{sec:app2}
	
	In this article, we employ a simplified, minimal approach to calculate the trajectory of the cargo particle in the flow created by the roller. Specifically, we calculate the flow field in the absence of any cargo particle using a finite-element routine, and then assume that the cargo simply follows the streamlines of this flow (except when altered by steric interactions). We employ this methodology since dynamic simulations in this geometry require computation times on the order of weeks to months for each individual data point, which is prohibitively expensive. However, it is of course still necessary to quantify the error incurred by this simplification.
	
	In order to assert the accuracy of our analysis, we calculate the flow field in the presence of a force- and torque-free cargo particle at six judiciously chosen positions and parameter configurations using the same finite-element routine with the remaining boundary conditions unchanged. These parameter configurations are listed on the left side of Table~\ref{tab:parameters}. Here, (a) to (d) are chosen to match the data points analysed in detail in Fig.~5 of the main text, while (e) and (f) represent an extreme point in the top right of the phase diagram of Fig.~4 of the main text, where we expect our analysis to be least accurate. We assume in each case that the cargo follows streamlines according to our minimal model up to the location where we calculate the flow field exactly. The value of $l/a$ is therefore to be understood as an identifier of the minimal model configuration that informs a particular cargo location, rather than a direct input to the numerical procedure.
	
	\begin{table}[t]
		\centering
		\begin{tabular}{c|c|c|c|c|c||c|c}
			& $b/a$ & $r/a$ & $l/a$ (see text) & Trapped? & Cargo location & $\Delta^2$ & $\beta$\\
			\hline
			(a) & 0.5 & 0.05 & 0.07 & No & Centre of squeezing & 0.1300 & $0.14^\circ$ \\
			(b) & 0.5 & 0.10 & 0.07 & Yes & Centre of squeezing & 0.1203 & $0.03^\circ$\\
			(c) & 0.5 & 0.10 & 0.07 & Yes & Halfway up vortex & 0.0033 &$0.16^\circ$ \\
			(d) & 0.5 & 0.25 & 0.07 & No & Centre of squeezing & 0.0179 & $0.10^\circ$\\
			(e) & 0.125 & 0.25 & 0.15 & Yes & Centre of squeezing & 0.0818 & $0.04^\circ$\\
			(f) & 0.125 & 0.25 & 0.15 & Yes & Halfway up vortex & 0.0033 & $0.67^\circ$\\
		\end{tabular}
		\caption{Parameter configurations for the verification of our methodology (left of double line) and two measures for the accuracy of the computed cargo velocity (right of double line).}\label{tab:parameters}
	\end{table}
	
	\begin{figure}[p!]
		\centering
		\begin{subfigure}[b]{.45\linewidth}
			\centering
			\includegraphics[width=.99\textwidth]{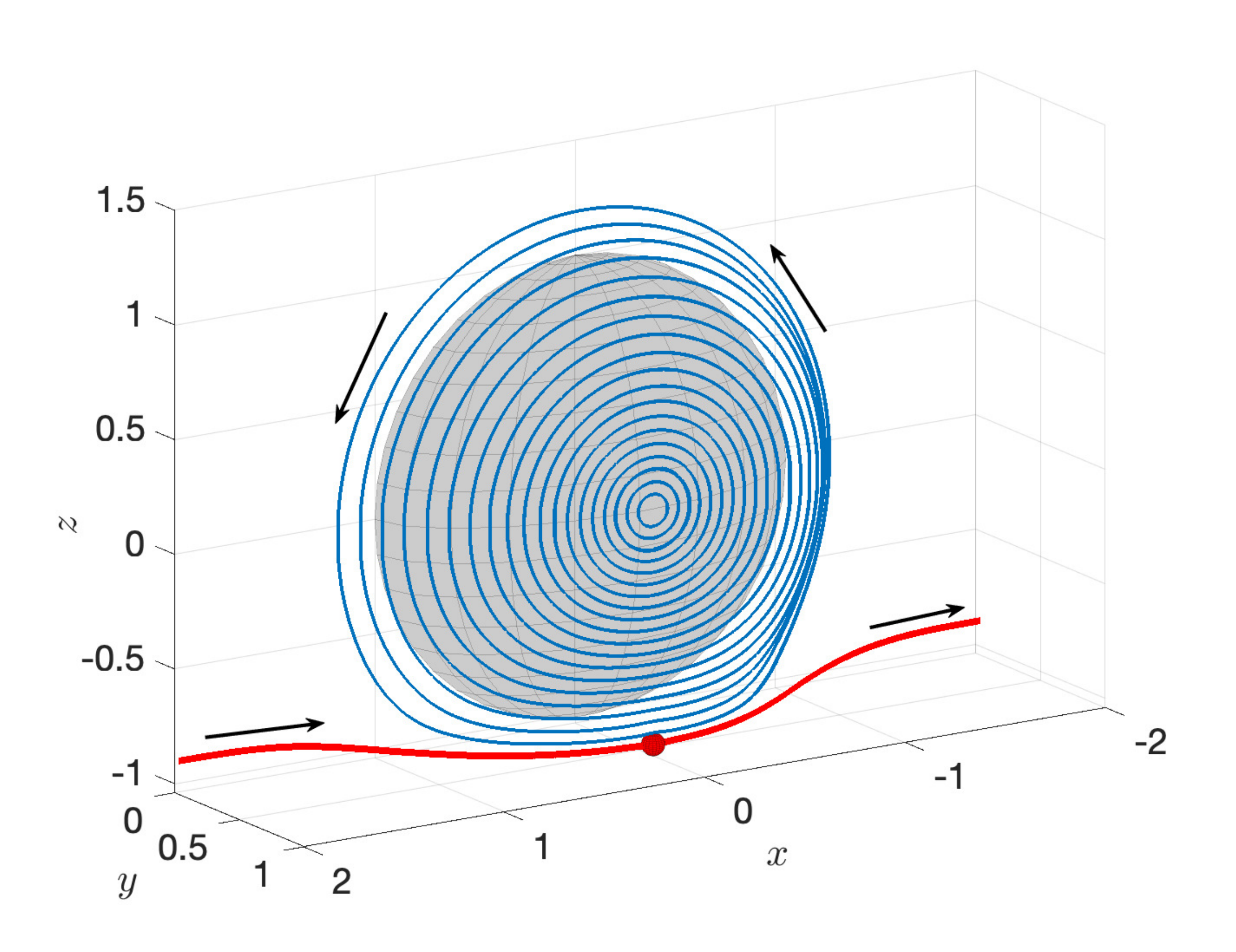}
			\caption{}
		\end{subfigure}
		\begin{subfigure}[b]{.45\linewidth}
			\centering
			\includegraphics[width=.99\textwidth]{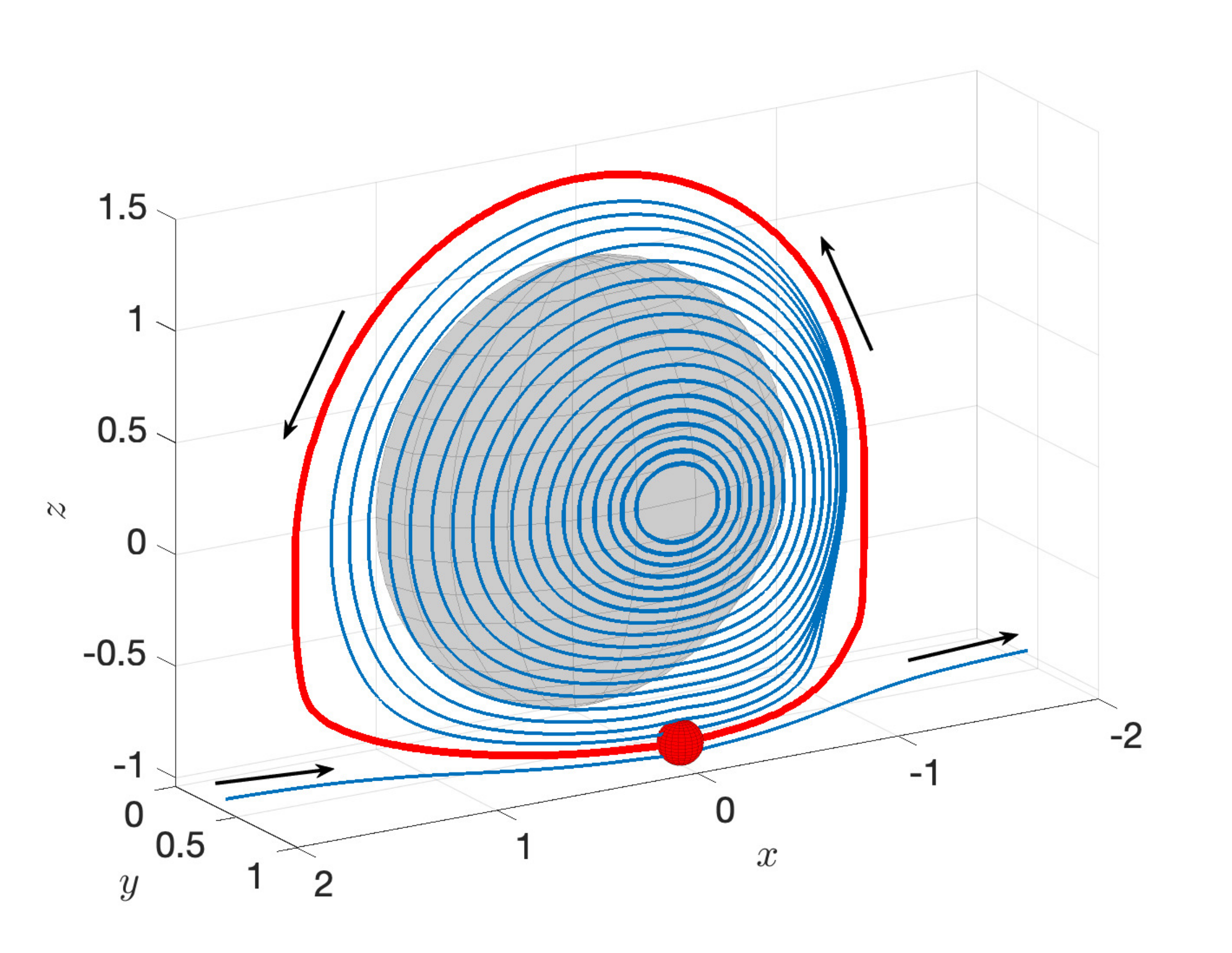}
			\caption{}
		\end{subfigure}
		\begin{subfigure}[b]{.45\linewidth}
			\centering
			\includegraphics[width=.99\textwidth]{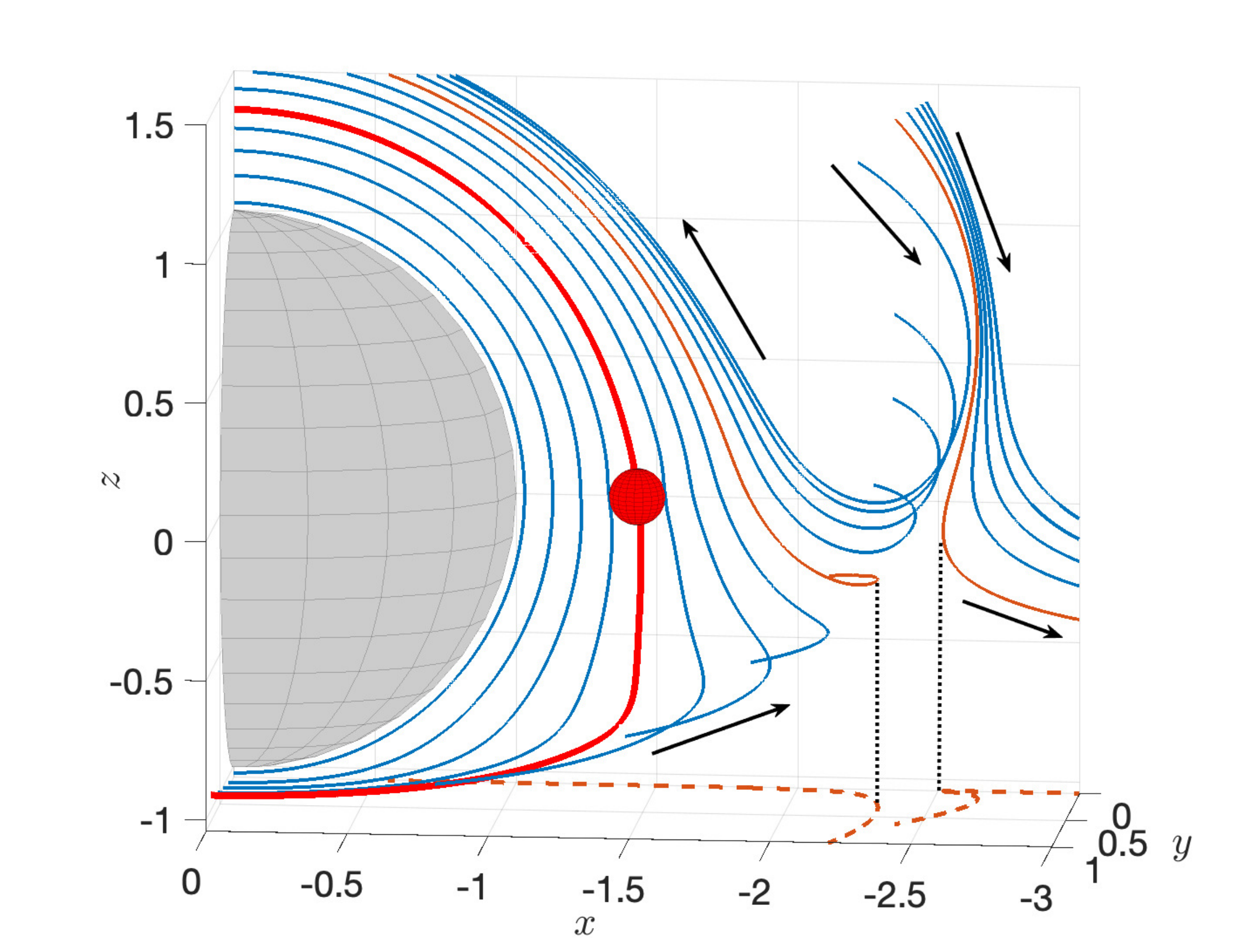}
			\caption{}
		\end{subfigure}
		\begin{subfigure}[b]{.45\linewidth}
			\centering
			\includegraphics[width=.99\textwidth]{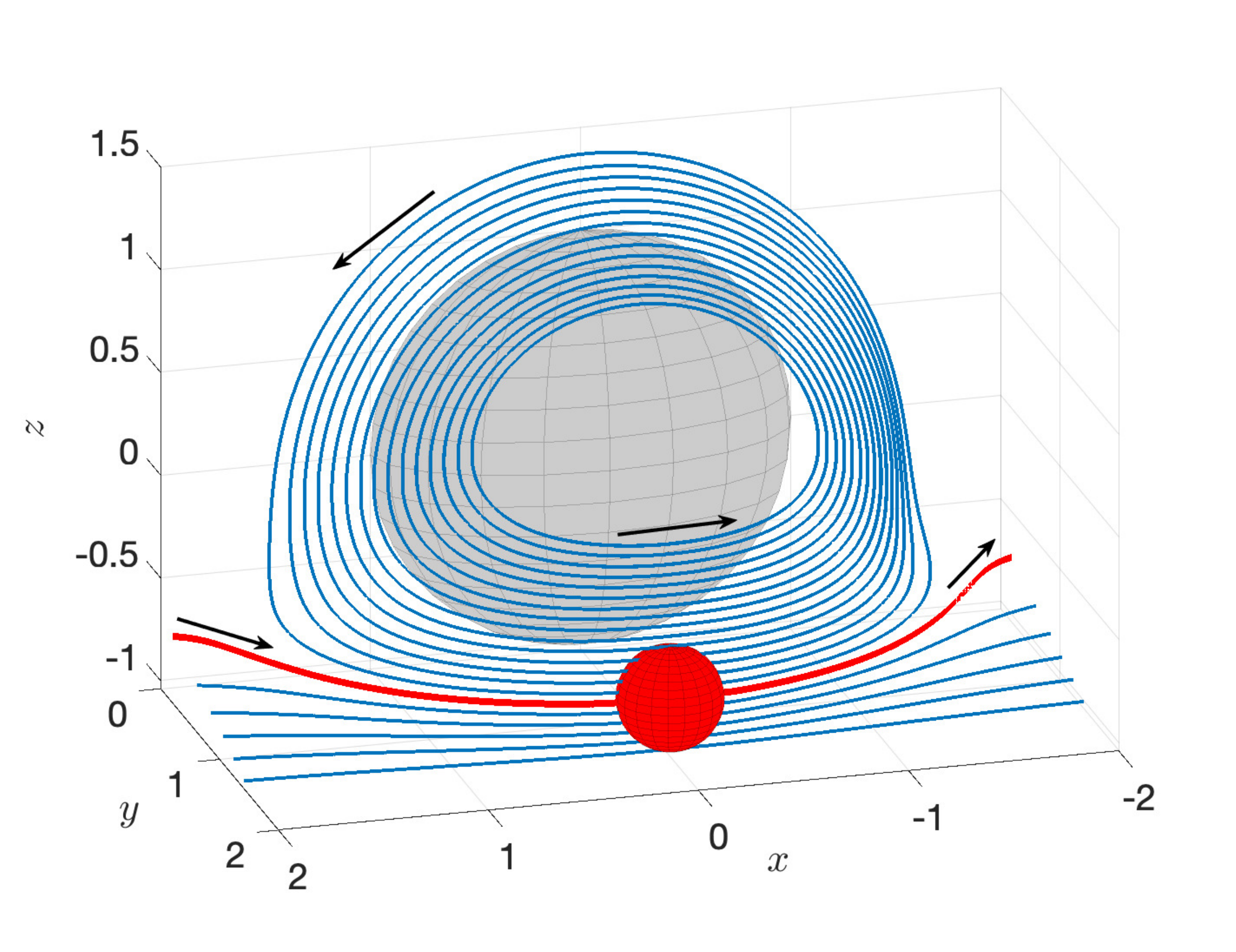}
			\caption{}
		\end{subfigure}
		\begin{subfigure}[b]{.45\linewidth}
			\centering
			\includegraphics[width=.99\textwidth]{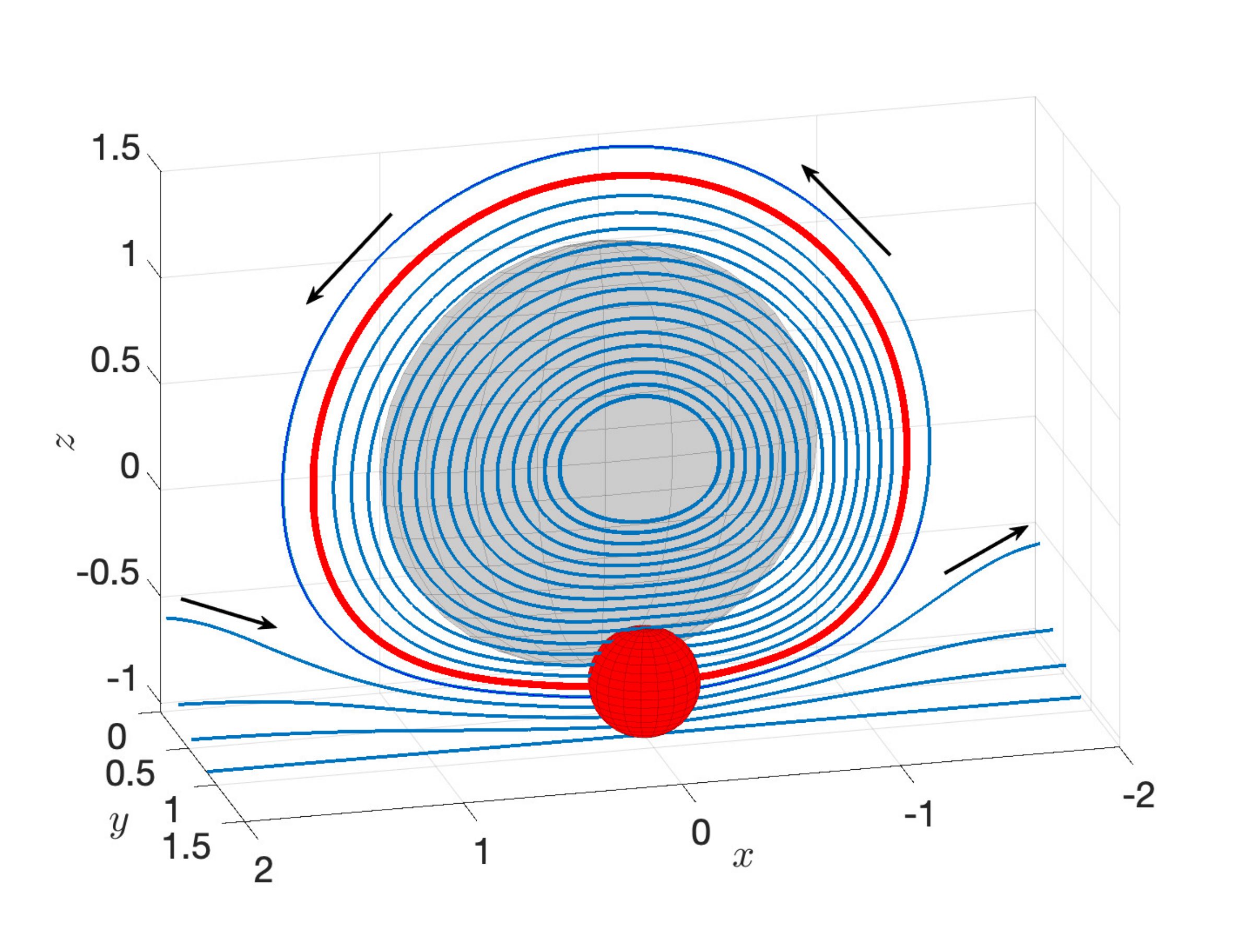}
			\caption{}
		\end{subfigure}
		\begin{subfigure}[b]{.45\linewidth}
			\centering
			\includegraphics[width=.99\textwidth]{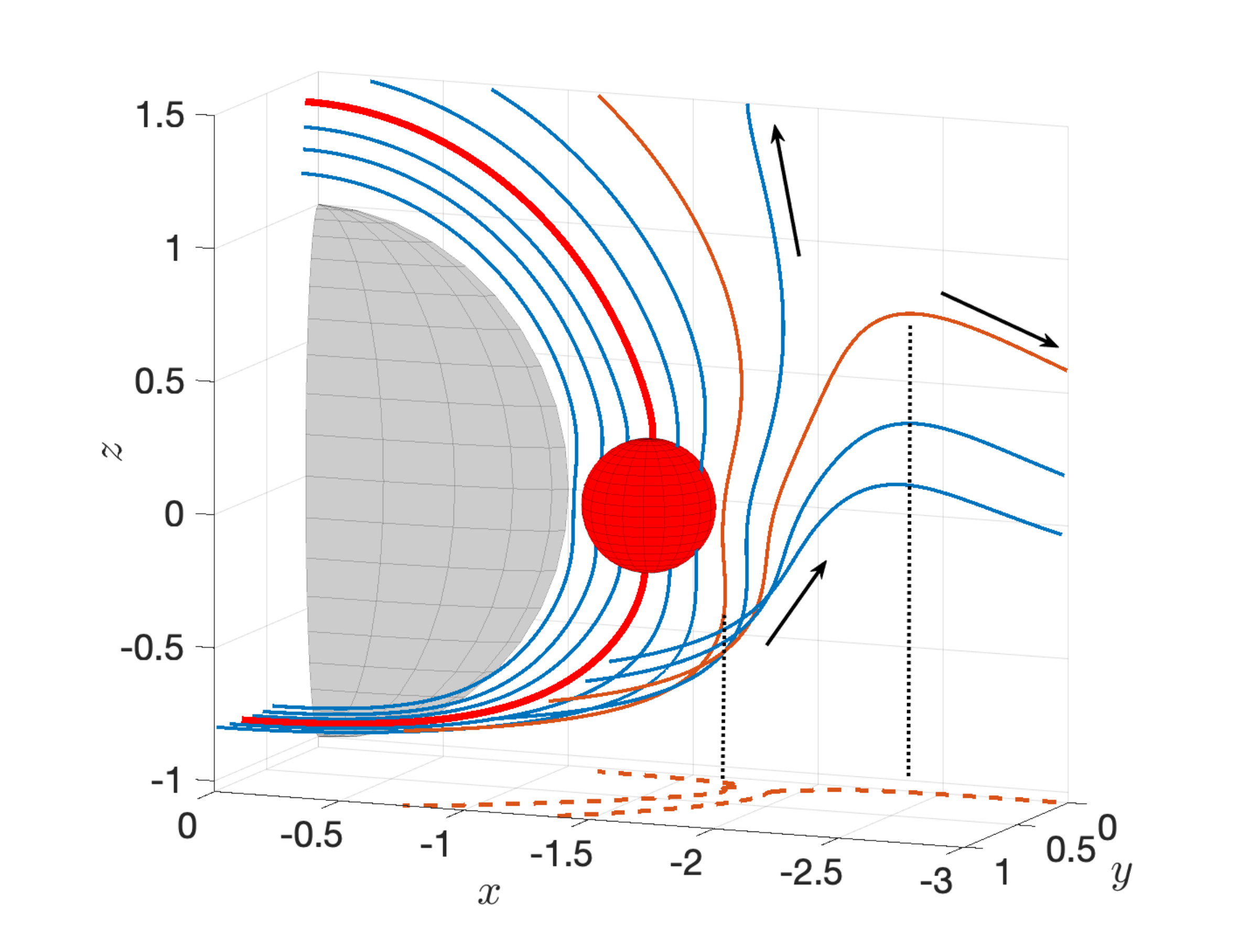}
			\caption{}
		\end{subfigure}
		\caption{Numerical illustration of streamlines for each of the parameter configurations listed in Table~\ref{tab:parameters}. The streamline through the cargo centre is highlighted in bold red, black arrows give an indication of flow direction. In (c) and (f) the shadows of certain (orange) streamlines are drawn in dashed orange to help visualise the flow topology.}
		\label{fig:verification}
	\end{figure}
	
	As illustrated in Fig.~\ref{fig:verification}, the nature of the streamline through the cargo centre is the same as in the simplified case presented in the main text in all cases considered. Furthermore, the qualitative disturbance of the flow topology due to the presence of the cargo is very small, even for a comparatively large particle with $r/a=0.25$ as shown in Fig.~\ref{fig:verification} (d) to (f). In the cases where the cargo is located at the centre of squeezing there is no visible deformation, while in the cases where the cargo is transported upward in the vortex there is a slight bending of the streamlines immediately in contact with the particle. This is due to a rigid-body rotation that the cargo experiences due to a non-zero vorticity of the flow. Crucially, the topology of the vortex remains intact. Even in the extreme case (f) the cargo particle is located well inside the vortex, with streamlines that escape to infinity separated from the cargo surface by more one cargo radius. However, as is illustrated in case (d), the picture is less clear at the centre of squeezing, where a slight upward dislocation of the cargo might lead to trapping. This threshold may conceivably be crossed even just due to thermal noise. The boundaries of our phase diagram may therefore be slightly blurred in a real system. Nevertheless, these results provide strong evidence that our methodology classifies particle trajectories accurately.
	
	To provide further quantitative evidence for the accuracy  of the methodology, we compute the velocity of the cargo particle $\bm{u}_c$ in the cases (a) to (f) and compare it with the flow velocity $\bm{u}(\bm{x}_c)$ at the position of the cargo centre $\bm{x}_c$ calculated in the absence of the particle. Since the velocities are vectors, we compare both the normalised squared difference in magnitude $\Delta^2=|\bm{u}_c-\bm{u}(\bm{x}_c)|^2/|\bm{u}_c|^2$, and the angle $\beta=\cos^{-1}\left(\bm{u}_c\cdot\bm{u}(\bm{x}_c)/|\bm{u}_c||\bm{u}(\bm{x}_c)|\right)$ between the velocity vectors. These results are summarised on the right side of Table~\ref{tab:parameters}. In all cases, the difference in direction is vanishingly small and amounts to less than $1^\circ$. The difference in magnitude is larger, especially when the cargo is located at the centre of squeezing. This is due to friction forces that could be calculated using lubrication theory. These are most significant in the cases (a) and (b) when the cargo is squeezed below the vortex while still passing close to the side of the roller, and smaller when the particle is deflected further to the side in the cases (d) and (e) and especially when it is located further away from rigid boundaries as in cases (c) and (f). Overall however the error remains small, and supports the modelling approach in the main text.

	\section{Derivation of the streamfunction for a rotating rigid disc}\label{sec:app3}
	
	The first important feature of the phase diagram is the prominence of trapping for rollers with a narrow aspect ratio. In order to elucidate this further, we begin by considering the extreme case of a rolling disc, i.e.~we consider the limit $b=0$, and in order to make analytical progress, we ignore the presence of the wall. We consider a frame in which the disc is stationary but rotating with angular velocity $\bm{\Omega}=\Omega\hat{\bm{y}}$, and scale lengths by the disc radius, $a$. Since the no-slip condition is applied on the disc's surface, very near to it (that is for $|y|$ small) the fluid is approximately in solid body rotation. In terms of cylindrical polar coordinates $(\rho,\theta,y)$ with $\rho^2=x^2+z^2$ and $\tan\theta=x/z$ we therefore seek a solution to the Stokes equations with boundary condition
	\begin{equation}\label{eq:BCpolar}
	\bm{u}=\Omega \rho\bm{e}_\theta,\quad y=0, \rho<1,
	\end{equation}
	and flow decaying to zero at infinity. For convenience, we introduce oblate spheroidal coordinates $(\lambda,\xi,\theta)$ defined by
	\begin{align}
	y &=\lambda\xi,\\
	\rho^2 &=(\lambda^2+1)(1-\xi^2),\\
	\theta &=\theta.
	\end{align}
	Note that $\lambda$ and $\theta$ are dimensionless, while $\xi$ has units of length. Surfaces of constant $\lambda$ are oblate spheroids that are defined by the relation
	\begin{equation}\label{eq:rholambda}
	\frac{\rho^2}{1+\lambda^2}+\frac{y^2}{\lambda^2}=1.
	\end{equation}
	In particular, the degenerate case $\lambda=0$ corresponds to a disc of radius $1$. Casting the problem in these coordinates therefore lends itself to a particularly convenient form of the boundary condition Eq.~\eqref{eq:BCpolar}, namely
	\begin{align}
	\bm{u}=\Omega \rho\bm{e}_\theta,\quad \lambda =0.
	\end{align}
	where $\rho(\lambda,\xi)$ is defined implicitly. It can be shown \cite{tanzosh1996general} that the solution is a purely azimuthal flow given 
	\begin{equation}
	\bm{u}=u_\theta(\rho,\lambda)\bm{e}_\theta,\quad u_\theta=\Omega \rho\times \frac{2}{\pi}\left(\cot^{-1}\lambda-\frac{\lambda}{1+\lambda^2}\right),
	\end{equation}
	which we can evaulate by using the relation
	\begin{equation}
	\lambda=\left\{{\frac{1}{2} \left(\rho^2+y^2-1\right)+\frac{1}{2} \left[{\left(\rho^2+y^2-1\right)^2+4
			y^2}\right]^{1/2}}\right\}^{1/2}.
	\end{equation}
	Since the flow is purely azimuthal in the $x$-$z$ plane, and therefore two-dimensional (2D) incompressible, we can define a streamfunction of the form $\bm{\psi}=\psi(\rho;y)\bm{\hat{y}}$ that recovers this flow field if we treat $y$ as a parameter that labels different `slices' of the fluid. Since $u_\theta=-\partial \psi/\partial \rho$ we have
	\begin{align}\label{eq:psidefint}
	\psi= \frac{2\Omega}{\pi}\int \left(\frac{\lambda}{1+\lambda^2}-\cot^{-1}\lambda \right)\rho\, \text{d}\rho,
	\end{align}
	We note that Eq.~\eqref{eq:rholambda} implies $\rho d\rho/d\lambda=\lambda+y^2/\lambda^3$ and so we can integrate Eq.~\eqref{eq:psidefint} exactly to find
	\begin{align}
	\psi&=\frac{\Omega}{\pi}\left[-3\frac{y^2}{\lambda}+\lambda+\left(\frac{y^2}{\lambda^2}+1+3 y^2- \lambda^2\right) \cot ^{-1}\lambda\right],
	\end{align}	
	where we choose the constant of integration such that $\psi\to0$ as $\lambda\to\infty$. This is the streamfunction for a rotating rigid disc in a quiescent infinite fluid.
	
}

\section{Details of the 2D singularity model}\label{sec:app4}

\subsection{Derivation}

\begin{figure}[h]
	\centering
	\includegraphics[width=0.5\textwidth]{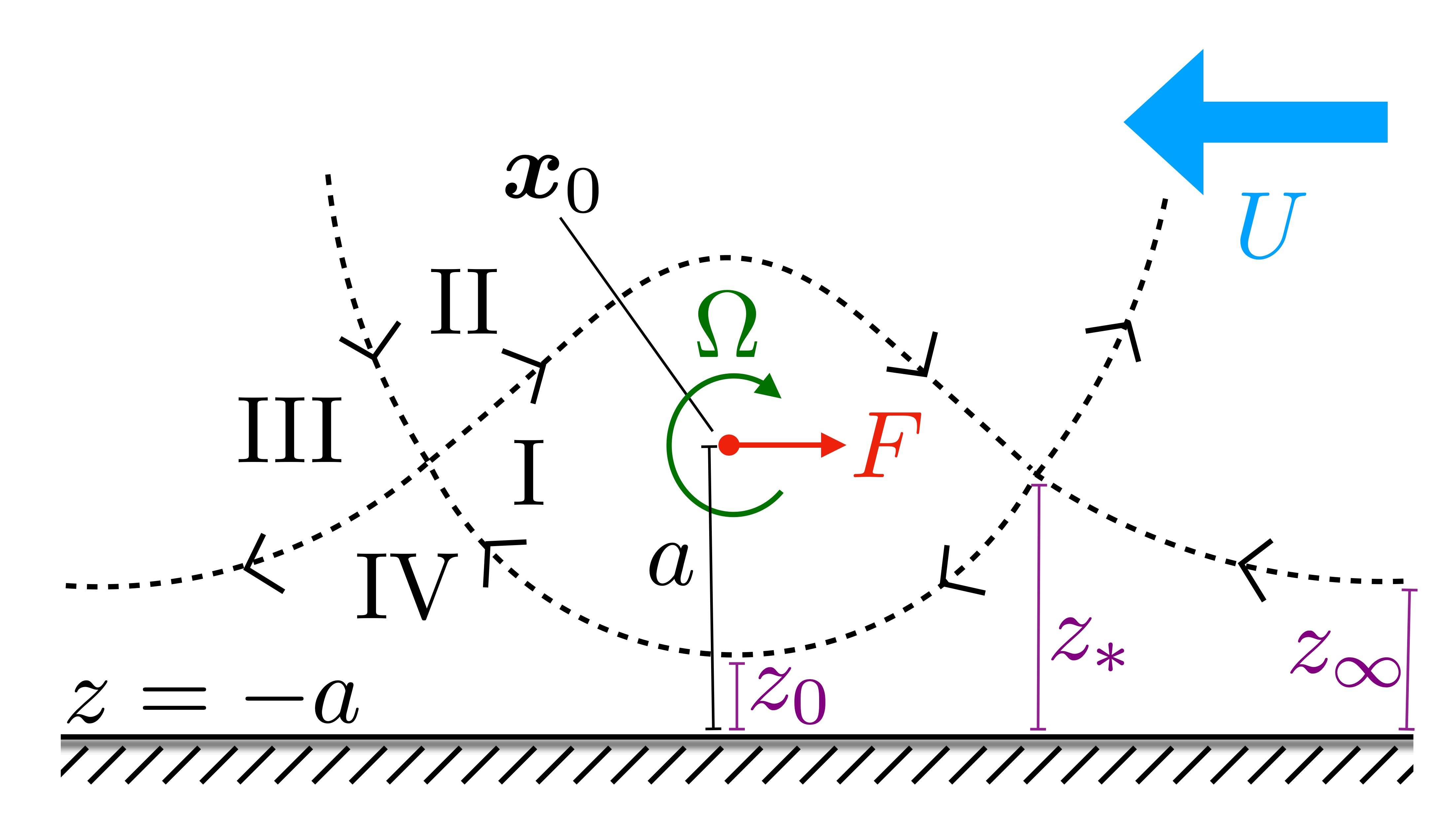}
	\caption{Sketch of the model geometry with the separatrix streamline and stagnation points.}\label{fig:sketchSI}
\end{figure}

Here we give some additional details for the 2D singularity model. We reproduce the sketch of the 2D singularity model in Fig.~\ref{fig:sketchSI}.

The Oseen tensor for 2D Stokes flow in the $x$-$z$ plane is given by\cite{pozrikidis1992boundary}
\begin{equation}
\bm{j}(\bm{x};\bm{x}_0)=-\log r \bm{I}+\frac{\bm{r}\bm{r}}{r^2},
\end{equation}
where $\bm{r}=\bm{x}-\bm{x}_0$ and $r=|\bm{r}|$. Some relevant derivatives are given by
\begin{align}
\partial_kj_{ij}&=\frac{-r_k\delta_{ij}+r_j\delta_{ik}+r_i\delta_{jk}}{r^2}-2\frac{r_ir_jr_k}{r^4},\quad&\text{(force dipole)}\\
\nabla^2j_{ij}&=2\frac{\delta_{ij}}{r^2}-4\frac{r_ir_j}{r^4}.\quad&\text{(source dipole)}
\end{align}
We consider the flow due to point singularities located at $\bm{x}_0=(0,0)$ in the presence of a rigid wall at $z=-a$ with normal $\bm{n}=(0,1)$. The flow due to a point force per unit length $\bm{F}$ is given in this geometry by
\begin{equation}\label{eq:stokeslet}
\bm{u}^f=\frac{\bm{F}}{8\pi\mu}\cdot\left(\bm{j}-\bm{j}^*-2a\bm{D}\cdot\nabla(\bm{j}^*\cdot\bm{n})+a^2\bm{D}\cdot\nabla^2\bm{j}^*\right),
\end{equation}
where $\bm{D}=\bm{I}-2\bm{n\bm{n}}$ and $\bm{j}^*=\bm{j}(\bm{x};\bm{D}\cdot\bm{x}_0)$. This has exactly the same structure as a point force in 3D flow \cite{blake1974fundamental}, and by linearity the same holds true for any higher order singularities. For a force parallel to the wall in the positive $x$-direction the expression in Eq.~\eqref{eq:stokeslet} evaluates to
\begin{equation}
\bm{u}^f=\frac{F}{8\pi\mu}\colvec{-\log r+\log R +\frac{x^2}{r^2}-\frac{x^2}{R^2}-\frac{2a(z+a)}{R^2}+\frac{4ax^2(z+a)}{R^4}}{\frac{xz}{r^2}-\frac{xz}{R^2}+\frac{4ax(z+a)(z+2a)}{R^4}},
\end{equation}
where $r^2=x^2+z^2$, and $R^2=x^2+(z+2a)^2$. 2D Stokes flow is incompressible and thus admits a streamfunction $\psi$ such that $\bm{u}=(\psi_z,-\psi_x)$. For the force parallel to the wall we then have
\begin{equation}
\psi^f=\frac{F}{8\pi\mu}\left(z\log\frac{R}{r}+\frac{2a(z+a)(z+2a)}{R^2}\right).
\end{equation}
For the flow due to a point vortex in the $x-z$ plane we consider the addition of a hypothetical $y$-axis with $\bm\Omega=(0,a\Omega,0)$ oriented along that axis. With this setup $\Omega>0$ corresponds to a clockwise rotation in the $x-z$ plane and thus rolling in the positive $x$-direction. The flow is given by
\begin{equation}\label{eq:rotlet}
\bm{u}^r=-\frac{1}{2}(\bm{\Omega}\times\nabla)\cdot\bm{j}+\frac{1}{2}(\bm{\Omega}\times\nabla)\cdot\bm{j}^*-\left(\bm{n} \times\bm\Omega\, \bm n+ \bm n \, \bm{n} \times\bm\Omega \right):\nabla\bm{j}^*+h(\bm{n} \times\bm\Omega)\cdot\nabla^2\bm{j}^*.
\end{equation}
This evaluates to the following flow in the $x$-$z$ plane:
\begin{equation}
\bm{u}^r=\Omega a \colvec{\frac{z}{r^2}-\frac{z}{R^2}+\frac{4x^2(z+a)}{R^4}}{-\frac{x}{r^2}+\frac{x}{R^2}+\frac{4x(z+a)(z+2a)}{R^4}}.
\end{equation}
The corresponding streamfunction is
\begin{equation}
\psi^r=\Omega a\left(-\log\frac{R}{r}+\frac{2(z+a)(z+2a)}{R^2}\right),
\end{equation}
which is actually quite similar to $\psi^f$ since similar image singularities are required for this solution. Finally we note that the streamfunction for a constant background flow in the negative $x$-direction $\bm{u}^b=(-U,0)$ is
\begin{equation}
\psi^b=-U(z+a).
\end{equation}
Most of these results have been derived previously, e.g.\ in \cite{Jeffrey1981}. All streamfunctions are defined so that they satisfy $\psi=0$ on the wall.

We scale lengths by $a=1$ from this point onwards and furthermore introduce the parameters $\eta=F/8\pi\mu a\Omega$ and $\gamma=U/a\Omega$ denoting the relative strength of the various terms. The combined and rescaled streamfunction is then
\begin{equation}
\psi=\left(\eta z-1\right)\log\frac{R}{r}+\frac{2(1+\eta)(z+1)(z+2)}{R^2}-\gamma (z+1),
\end{equation}
as claimed in the main text. In the following, we analyse the stagnation points and topology of the streamlines in the two cases $\eta=0$ (force-free) and $\eta>0$ (with force).

\begin{figure}[h]
	\centering
	\begin{subfigure}[b]{0.3\textwidth}
		\includegraphics[width=\textwidth]{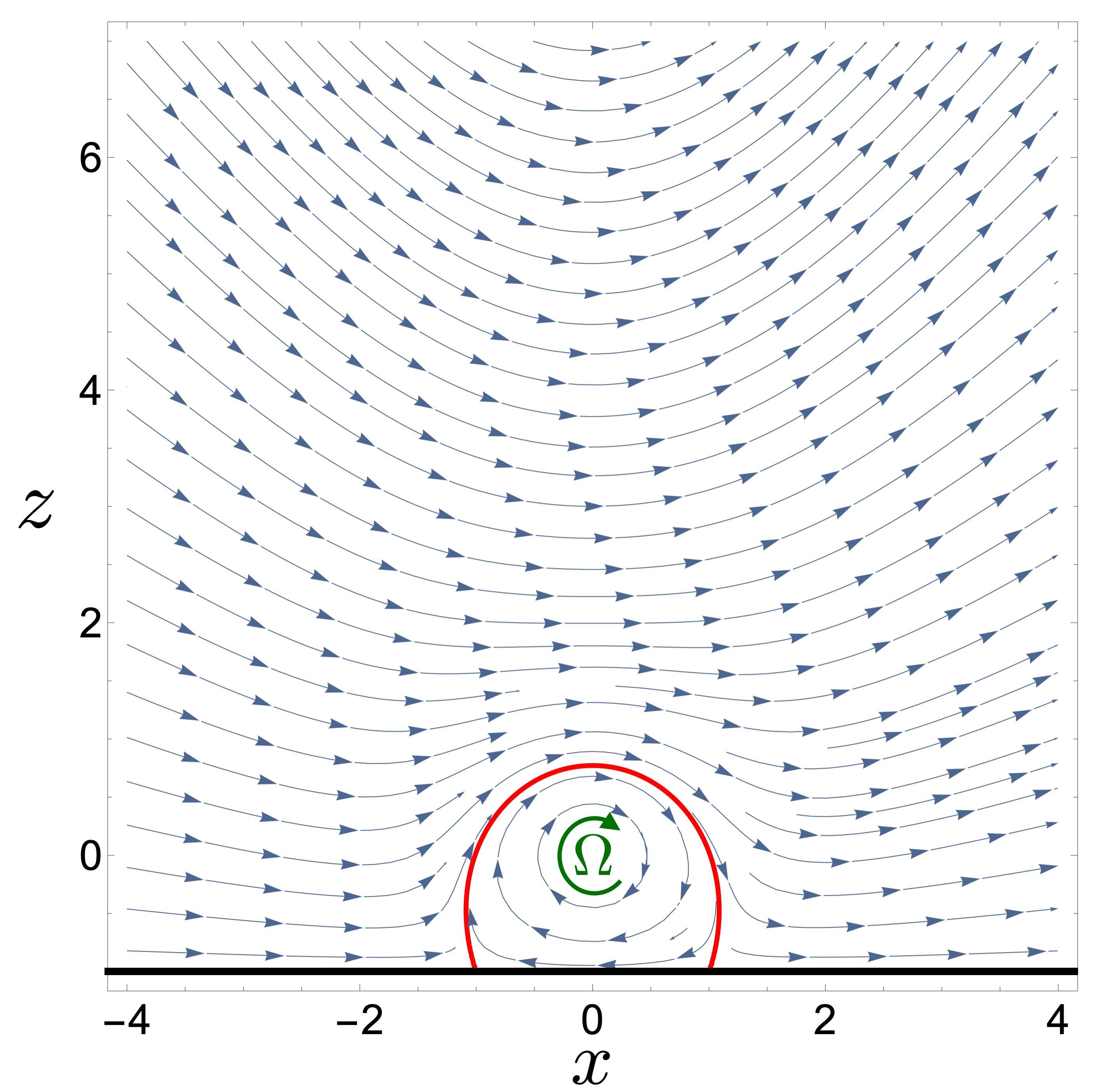}
		\caption{$\gamma=0$}
	\end{subfigure}
	\begin{subfigure}[b]{0.3\textwidth}
		\includegraphics[width=\textwidth]{G025.jpg}
		\caption{$\gamma=0.25$}
	\end{subfigure}
	\begin{subfigure}[b]{0.3\textwidth}
		\includegraphics[width=\textwidth]{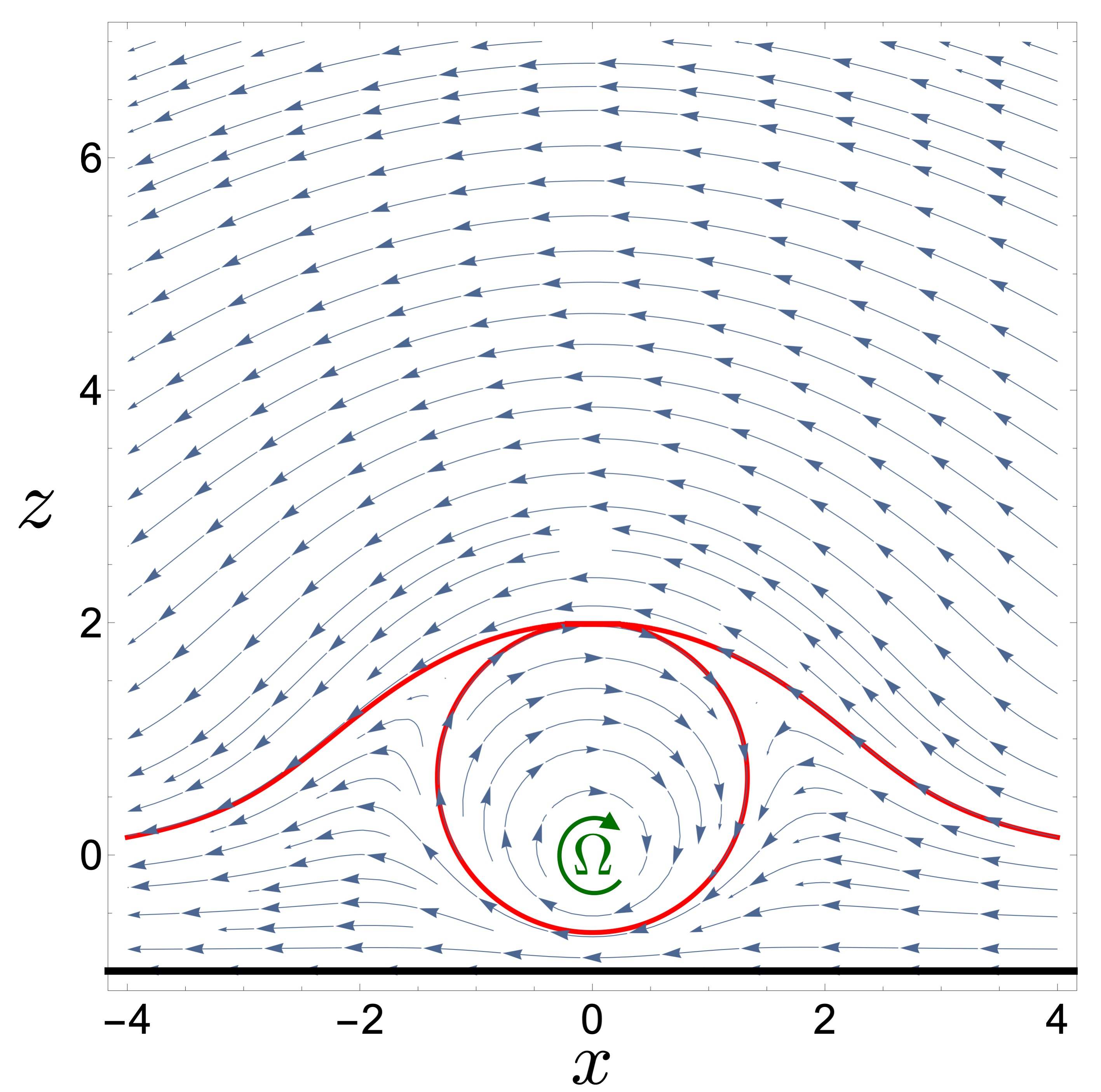}
		\caption{$\gamma=3/8$}
	\end{subfigure}
	\caption{Streamlines for $\eta=0$ ((no force). The separatrix streamline is indicated in bold red. For $\gamma>0$ we see that it is squeezed below the singularity.}\label{fig:nf}
\end{figure}

\subsection{No force, $\eta=0$}
Let us first consider the case of a force-free roller, i.e.\ $\eta=0$. Some sample streamlines are plotted in Figure \ref{fig:nf}. Our first goal is to find the stagnation points of the flow. Upon differentiating $\psi$ with respect to $x$ we find that the vertical velocity is zero when $z=-1$, $x=0$ or when the condition
\begin{equation}
x^2+z^2=4
\end{equation}
is satisfied. Differentiating $\psi$ with respect to $z$ and substituting for $x$ we can determine the position of the stagnation points exactly and find that
\begin{align}
\nabla\psi=\bm{0}\quad\text{if}\quad(x,z)&=\left(\pm\frac{\sqrt{3-8\gamma}}{1-2\gamma},\frac{4\gamma-1}{1-2\gamma}\right),\quad 0\leq\gamma<\frac{3}{8},\\
\text{or}\quad(x,z)&=(0,z'),
\end{align}
where $z'$ is solution to
\begin{align}
\frac{4(1+z')}{z'(2+z')^2}=\gamma.
\end{align}
For $0<\gamma<3/8$ the first two constitute saddle points fore and aft the roller, while the third corresponds to a centre vertically above the singularity. When $\gamma=0$, the centre disappears and the saddle points collapse onto the wall. When $\gamma$ passes through $3/8$ then these coalesce in a pitchfork bifurcation and only one saddle remains. Since we observe $\gamma\approx0.1$ in our numerical simulations, we discard this case and obtain a flow field with four topologically distinct regions as discussed in the main text.

The value of the streamfunction at the stagnation points is
\begin{equation}
\psi_0=\gamma+\frac{1}{2}\log(1-2\gamma)=-\gamma^2-\frac{4}{3}\gamma^3+\mathcal{O}(\gamma^4),
\end{equation}
so that the streamline passing through the stagnation point satisfies $\psi=\psi_0$. We compare the height of this streamline centrally below the singularity ($z_0$), at the stagnation point ($z_*$) and far away ($z_\infty$) to understand whether squeezing occurs. As $x\to\infty$ we have
\begin{equation}
\psi=-\gamma (z+1)+2\frac{(z+1)^2}{x^2}+\mathcal{O}(x^{-4}),\quad\Rightarrow\quad z_\infty=-1-\frac{1}{2\gamma}\log\left(1-2\gamma\right).
\end{equation} 
At leading order in $\gamma$ we therefore have $z_\infty=\gamma$. Meanwhile, $z_0$ satisfies
\begin{equation}
\gamma+\frac{1}{2}\log(1-2\gamma)=\log\frac{1-z_0}{1+z_0}+\frac{2z_0}{1+z_0}-\gamma z_0.
\end{equation}
Due to the presence of the logarithm, this equation does not have an analytic solution. However, we may expand for small $z_0$ and $\gamma$ and find
\begin{equation}
-\gamma^2-\frac{4}{3}\gamma^3+\dots=-\gamma z_0-2z_0^2+\frac{4}{3}z_0^3+\dots
\end{equation}
Thus $z_0\sim \gamma$ at leading order and we can solve a quadratic to find $z_0=\gamma/2+O(\gamma^2)$. In summary we have to leading order that
\begin{equation}
z_0=\frac{1}{2}\gamma,\quad z_*=2\gamma,\quad z_\infty=\gamma,\quad \psi_0=-\gamma^2.
\end{equation}
Thus the streamline coming in from infinity first goes up to twice its original height at the stagnation point before being squeezed down to half its original height below the singularity. As discussed in the main text, this squeezing of streamlines gives rise to irreversible trapping of cargo particles. An illustration is given in Figure \ref{fig:nf}.

\begin{figure}[t!]
	\centering
	\begin{subfigure}[b]{0.49\textwidth}
		\includegraphics[width=\textwidth]{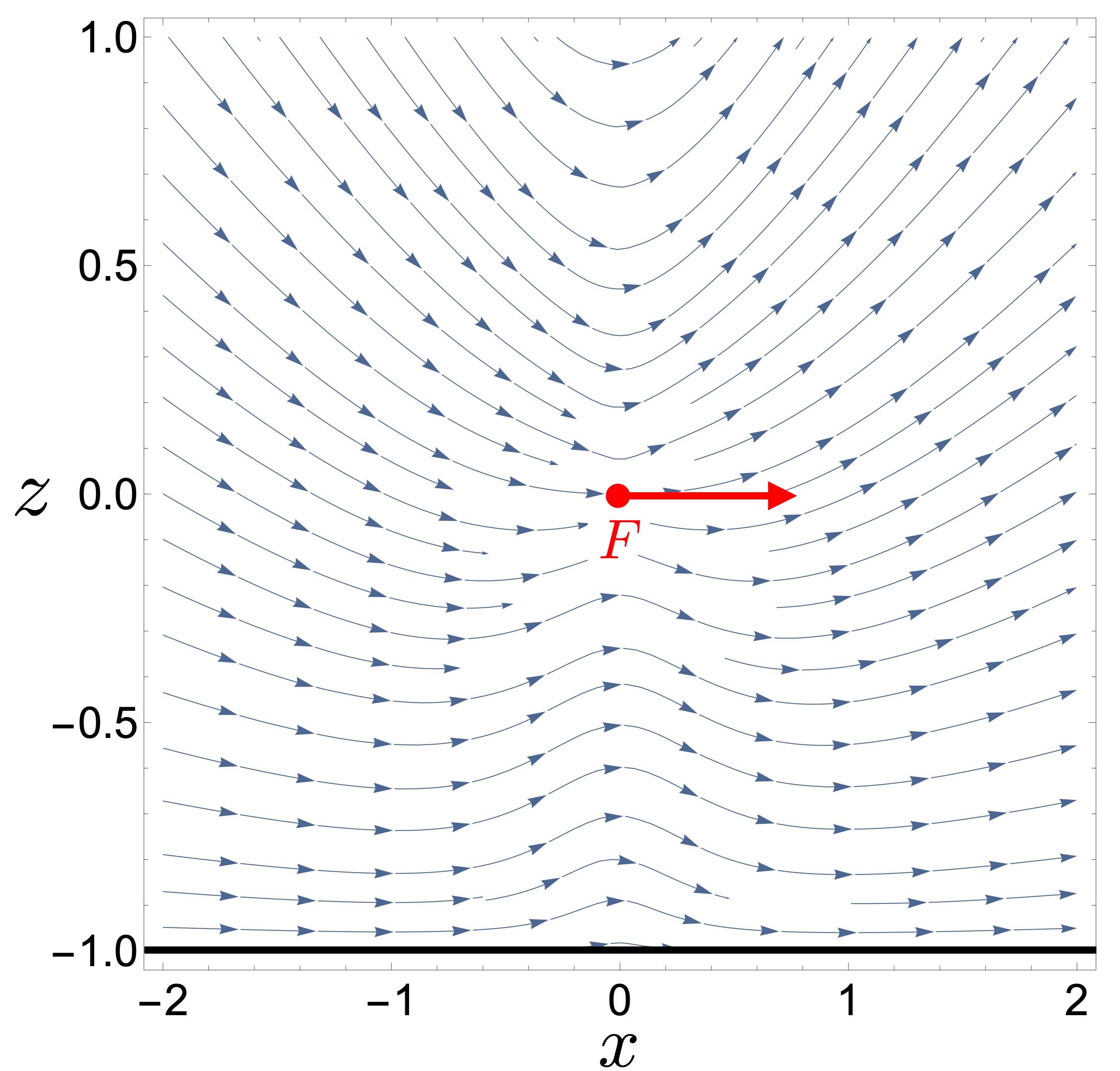}
		\caption{$\gamma/\eta=0$}
	\end{subfigure}
	\begin{subfigure}[b]{0.49\textwidth}
		\includegraphics[width=\textwidth]{F05.jpg}
		\caption{$\gamma/\eta=0.5$}
	\end{subfigure}
	\caption{Streamlines for a force with no rotation ($\psi/\eta$ for $\eta\to\infty$ and $\gamma/\eta$ finite). The separatrix streamline is indicated in bold red. No squeezing occurs, yet a region of closed streamlines exists below the singularity.}\label{fig:ofSI}
\end{figure}

\subsection{With force, $\eta>0$}
In the case that the force is non-zero we have a more complicated streamfunction. 
In this case the condition for no vertical flow ($\psi_x=0$) becomes
\begin{equation}
x^2+\left(z+\frac{2\eta}{1+2\eta}\right)^2=4\left(\frac{1+\eta}{1+2\eta}\right)^2.
\end{equation}
Using this we find that the condition for no lateral velocity ($\psi_z=0$) becomes
\begin{equation}
-\gamma+\frac{(1+2\eta)(z+1)}{2(z+2)}+\frac{\eta}{2}\log\left(\frac{(1+\eta)(z+2)}{1-\eta z}\right)=0.
\end{equation}
This is now a transcendental equation for $z$ with no analytical solution. To make progress, we expand this for small $\gamma$ and find that
\begin{equation}
z_*=\frac{2(1+\eta)}{(1+2\eta)^2}\gamma+\mathcal{O}(\gamma^2),\quad\psi_0=-\frac{1+\eta}{(1+2\eta)^2}\gamma^2+\mathcal{O}(\gamma^3).
\end{equation}
For as $x\to\infty$ we have
\begin{equation}
\psi=-\gamma (z+1) +\frac{(2+4\eta)(z+1)^2}{x^2}+\mathcal{O}(x^{-4}),
\end{equation}
so that $z_\infty=(1+\eta)\gamma/(1+2\eta)^2+\mathcal{O}(\gamma^2)$. For $z_0$ we find
\begin{equation}
-\frac{1+\eta}{(1+2\eta)^2}\gamma^2+\mathcal{O}(\gamma^3)=-\gamma z_0 - 2 z_0^2 + \frac{4(1+\eta)}{3} z_0^3+\mathcal{O}(z_0^4)
\end{equation}
so that to leading order $z_0=\gamma/2(1+2\eta)$. In summary,
\begin{equation}
z_0=\frac{1}{2+4\eta}\gamma,\quad z_*=\frac{2(1+\eta)}{(1+2\eta)^2}\gamma,\quad z_\infty=\frac{1+\eta}{(1+2\eta)^2}\gamma,\quad \psi_0=-\frac{1+\eta}{(1+2\eta)^2}\gamma^2,
\end{equation}
as quoted in the main text. As expected, we recover our previous results if we set $\eta=0$. We also have
\begin{equation}
\frac{z_0}{z_\infty}=\frac{1+2\eta}{2+2\eta}\leq 1
\end{equation}
for $\eta\geq 0$, therefore squeezing always occurs in the presence of a force. However, the relative extent to which streamlines are squeezed is maximised for a force-free roller. Notably, in the limit $\eta\to\infty$ with $\gamma/\eta$ finite, corresponding to a purely translating roller with no rotation we have
\begin{equation}
z_0=\frac{1}{4\eta}\gamma,\quad z_*=\frac{1}{2\eta}\gamma,\quad z_\infty=\frac{1}{4\eta}\gamma,\quad \psi_0=-\frac{1}{4\eta}\gamma^2,
\end{equation}
indicating that no squeezing occurs. This shows that according to our model rotation is an essential ingredient for entrapment. An illustration is given in Figure \ref{fig:ofSI}.

\begin{acknowledgments}
	S.T. is supported by a Nakajima Foundation Scholarship and a John Lawrence Cambridge Trust International Scholarship. This project has also received funding from the European Research Council (ERC) under the European Union's Horizon 2020 research and innovation programme  (grant agreement 682754 to E.L.).
\end{acknowledgments}

\bibliographystyle{abbrv}
\bibliography{references.bib}
\end{document}